\documentclass[draftclsnofoot,onecolumn,11pt,journal]{IEEEtran}

\usepackage{times}
\usepackage{epsfig}
\usepackage{graphicx}
\usepackage{amsmath}
\usepackage{amssymb}
\usepackage{amsthm}
\usepackage{enumerate}
\usepackage{indentfirst}
\usepackage{subfigure}
\usepackage{algorithmic}
\usepackage{algorithm}
\usepackage{url}
\usepackage{booktabs}
\usepackage{multirow}
\usepackage{longtable}
\usepackage{cite}

\theoremstyle{remark}
\newtheorem{theorem}{Theorem}
\newtheorem{proposition}{Proposition}
\newtheorem{lemma}{Lemma}

\long\def\symbolfootnote[#1]#2{\begingroup\def\thefootnote{\fnsymbol{footnote}}
\footnote[#1]{#2}\endgroup}

\ifCLASSINFOpdf
\else
\fi
\begin{document}
%
% paper title
% can use linebreaks \\ within to get better formatting as desired
\title{Performance Analysis of Heterogeneous Feedback Design in an OFDMA Downlink \\with Partial and Imperfect Feedback}
%
%
% author names and IEEE memberships
% note positions of commas and nonbreaking spaces ( ~ ) LaTeX will not break
% a structure at a ~ so this keeps an author's name from being broken across
% two lines.
% use \thanks{} to gain access to the first footnote area
% a separate \thanks must be used for each paragraph as LaTeX2e's \thanks
% was not built to handle multiple paragraphs
%

\author{Yichao Huang, \IEEEmembership{Member, IEEE}, and Bhaskar D. Rao, \IEEEmembership{Fellow,
IEEE}
%        Department of Electrical and Computer Engineering\\
%        University of California, San Diego\\
%        La Jolla, CA 92093-0407, USA\\
%        Email: \{yih006, brao\}@ucsd.edu
\thanks{Copyright (c) 2012 IEEE. Personal use of this material is permitted. However, permission to use this material for any other purposes must be obtained from the IEEE by sending a request to pubs-permissions@ieee.org.}
\thanks{This research was supported by Ericsson endowed chair funds, the Center for Wireless Communications, UC Discovery grant com09R-156561 and NSF
grant CCF-1115645. The material in this paper was presented in part at the 45th Asilomar Conference on Signals, Systems, and Computers, Pacific
Grove, CA, November 2011.}
\thanks{The authors are with Department of Electrical and Computer Engineering, University of California, San Diego, La Jolla, CA
92093-0407, USA (e-mail: yih006@ucsd.edu; brao@ece.ucsd.edu).}}

%\thanks{Manuscript received April 19, 2005; revised January 11, 2007.}}

% note the % following the last \IEEEmembership and also \thanks -
% these prevent an unwanted space from occurring between the last author name
% and the end of the author line. i.e., if you had this:
%
% \author{....lastname \thanks{...} \thanks{...} }
%                     ^------------^------------^----Do not want these spaces!
%
% a space would be appended to the last name and could cause every name on that
% line to be shifted left slightly. This is one of those "LaTeX things". For
% instance, "\textbf{A} \textbf{B}" will typeset as "A B" not "AB". To get
% "AB" then you have to do: "\textbf{A}\textbf{B}"
% \thanks is no different in this regard, so shield the last } of each \thanks
% that ends a line with a % and do not let a space in before the next \thanks.
% Spaces after \IEEEmembership other than the last one are OK (and needed) as
% you are supposed to have spaces between the names. For what it is worth,
% this is a minor point as most people would not even notice if the said evil
% space somehow managed to creep in.

% The paper headers
%\markboth{Journal of \LaTeX\ Class Files,~Vol.~6, No.~1, January~2007}%
\markboth{To Appear In IEEE Transactions on Signal Processing}%
{To Appear In IEEE Transactions on Signal Processing}
% The only time the second header will appear is for the odd numbered pages
% after the title page when using the twoside option.
%
% *** Note that you probably will NOT want to include the author's ***
% *** name in the headers of peer review papers.                   ***
% You can use \ifCLASSOPTIONpeerreview for conditional compilation here if
% you desire.

% If you want to put a publisher's ID mark on the page you can do it like
% this:
%\IEEEpubid{0000--0000/00\$00.00~\copyright~2007 IEEE}
% Remember, if you use this you must call \IEEEpubidadjcol in the second
% column for its text to clear the IEEEpubid mark.

% use for special paper notices
%\IEEEspecialpapernotice{(Invited Paper)}

% make the title area
\maketitle

\begin{abstract}
%\boldmath
Current OFDMA systems group resource blocks into subband to form the basic feedback unit. Homogeneous feedback design with a common subband size
is not aware of the heterogeneous channel statistics among users. Under a general correlated channel model, we demonstrate the gain of matching
the subband size to the underlying channel statistics motivating heterogeneous feedback design with different subband sizes and feedback
resources across clusters of users. Employing the best-M partial feedback strategy, users with smaller subband size would convey more partial
feedback to match the frequency selectivity. In order to develop an analytical framework to investigate the impact of partial feedback and
potential imperfections, we leverage the multi-cluster subband fading model. The perfect feedback scenario is thoroughly analyzed, and the
closed form expression for the average sum rate is derived for the heterogeneous partial feedback system. We proceed to examine the effect of
imperfections due to channel estimation error and feedback delay, which leads to additional consideration of system outage. Two transmission
strategies: the fix rate and the variable rate, are considered for the outage analysis. We also investigate how to adapt to the imperfections in
order to maximize the average goodput under heterogeneous partial feedback.
\end{abstract}
% IEEEtran.cls defaults to using nonbold math in the Abstract.
% This preserves the distinction between vectors and scalars. However,
% if the journal you are submitting to favors bold math in the abstract,
% then you can use LaTeX's standard command \boldmath at the very start
% of the abstract to achieve this. Many IEEE journals frown on math
% in the abstract anyway.

% Note that keywords are not normally used for peerreview papers.

% For peer review papers, you can put extra information on the cover
% page as needed:
% \ifCLASSOPTIONpeerreview
% \begin{center} \bfseries EDICS Category: 3-BBND \end{center}
% \fi
%
% For peerreview papers, this IEEEtran command inserts a page break and
% creates the second title. It will be ignored for other modes.
\IEEEpeerreviewmaketitle

\section{Introduction}\label{introduction}
% The very first letter is a 2 line initial drop letter followed
% by the rest of the first word in caps.
%
% form to use if the first word consists of a single letter:
% \IEEEPARstart{A}{demo} file is ....
%
% form to use if you need the single drop letter followed by
% normal text (unknown if ever used by IEEE):
% \IEEEPARstart{A}{}demo file is ....
%
% Some journals put the first two words in caps:
% \IEEEPARstart{T}{his demo} file is ....
%
% Here we have the typical use of a "T" for an initial drop letter
% and "HIS" in caps to complete the first word.
%\IEEEPARstart{T}{his} demo file is intended to serve as a ``starter file''
%for IEEE journal papers produced under \LaTeX\ using
%IEEEtran.cls version 1.7 and later.
% You must have at least 2 lines in the paragraph with the drop letter
% (should never be an issue)

% needed in second column of first page if using \IEEEpubid
%\IEEEpubidadjcol
Leveraging feedback to obtain the channel state information at the transmitter (CSIT) enables a wireless system to adapt its transmission
strategy to the varying wireless environment. The growing number of wireless users, as well as their increasing demands for higher data rate
services impose a significant burden on the feedback link. In particular for OFDMA systems, which have emerged as the core technology in 4G and
future wireless systems, full CSIT feedback may become prohibitive because of the large number of resource blocks. This motivates more efficient
feedback design approaches in order to achieve performance comparable to a full CSIT system with reduced feedback. In the recent years,
considerable work and effort has been focused on limited or partial feedback design, e.g., see \cite{love08} and the references therein. To the
best of our knowledge, most of the existing partial feedback designs are homogeneous, i.e., users' feedback consumptions do not adapt to the
underlying channel statistics. In this paper, we propose and analyze a heterogeneous feedback design, which aligns users' feedback needs to the
statistical properties of their wireless environments.

Current homogeneous feedback design in OFDMA systems groups the resource blocks into subband \cite{zhu09} which forms the basic scheduling and
feedback unit. Since the subband granularity is determined by the frequency selectivity, or the coherence bandwidth of the underlying channel,
it would be beneficial to adjust the subband size of different users according to their channel statistics. Empirical measurements and analysis
from the channel modeling field have shown that the root mean square (RMS) delay spread which is closely related to the coherence bandwidth, is
both location and environment dependent \cite{asplund06, huang12}. The typical RMS delay spread for an indoor environment in WLAN does not
exceed hundreds of nanoseconds; whereas in the outdoor environment of a cellular system, it can be up to several microseconds. Intuitively,
users with lower RMS delay spread could model their channel with a larger subband size and require less feedback resource than the users with
higher RMS delay spread. Herein, we investigate this heterogeneous feedback design in a multiuser opportunistic scheduling framework where the
system favors the user with the best channel condition to exploit multiuser diversity \cite{knopp95, viswanath02}. There are two major existing
partial feedback strategies for opportunistic scheduling, one is based on thresholding where each user provides one bit of feedback per subband
to indicate whether or not the particular channel gain exceeds a predetermined or optimized threshold \cite{sanayei07, hassel07, chen08,
pugh10}. The other promising strategy currently considered in practical systems such as LTE \cite{sesia11} is the best-M strategy, where the
receivers order and convey the M best channels \cite{jung07, ko07, choi07, choi08, pedersen09, leinonen09, donthi11, hur11}. The best-M partial
feedback strategy is embedded in the proposed heterogeneous feedback framework. Apart from the requirement of partial feedback to save feedback
resource, the study of imperfections is also important to understand the effect of channel estimation error and feedback delay on the
heterogeneous feedback framework. These imperfections are also considered in our work.

\subsection{Focus and Contributions of the Paper}
An important step towards heterogeneous feedback design is leveraging the ``match" among coherence bandwidth, subband size and partial feedback.
Under a given amount of partial feedback, if the subband size is much larger than the coherence bandwidth, then multiple independent channels
could exist within a subband and the subband-based feedback could only be a coarse representative of the channels. On the other hand, if the
subband size is much smaller than the coherence bandwidth, then channels in adjacent subbands are likely to be highly correlated and requiring
feedback on adjacent subbands could be a waste of resource; or a small amount of subband-based partial feedback may not be enough to reflect the
channel quality. In order to support this heterogeneous framework, we first consider the scenario of a general correlated channel model with one
cluster of users with the same coherence bandwidth. The subband size is adjustable and each user employs the best-M partial feedback strategy to
convey the M best channel quality information (CQI) which is defined to be the subband average rate. The simulation result  shows that a
suitable chosen subband size yields higher average sum rate under partial feedback conforming the aforementioned intuition. This motivates the
design of heterogeneous feedback to ``match" the subband size to the coherence bandwidth. The above-mentioned study, though closely reflects the
relevant mechanism, is not analytically tractable due to two main reasons. Firstly, the general correlated channel model complicates the
statistical analysis of the CQI. Secondly, the use of subband average rate as CQI makes it difficult to analyze the multi-cluster scenario.
Therefore, a simplified generic channel model is needed that balances the competing needs of analytical tractability and practical relevance.

In order to facilitate analysis, a subband fading channel model is developed that generalizes the widely used frequency domain block fading
channel model. The subband fading model is suited for the multi-cluster analysis. According to the subband fading model, the channel frequency
selectivity is flat within each subband, and independent across subbands. Since the subband sizes are different across different clusters, the
number of independent channels are heterogeneous across clusters and this yields heterogeneous partial feedback design. Another benefit of the
subband fading model is that the CQI becomes the channel gain and thus facilitate further statistical analysis. Under the multi-cluster subband
fading model\footnote[1]{An initial treatment of a two-cluster scenario was first presented in \cite{huang11}.} and the assumption of perfect
feedback, we derive a closed form expression for the average sum rate. Additionally, we approximate the sum rate ratio for heterogeneous design,
i.e., the ratio of the average sum rate obtained by a partial feedback scheme to that achieved by a full feedback scheme, in order to choose
different best-M for users with different coherence bandwidth. We also compare and demonstrate the potential of the proposed heterogeneous
feedback design against the homogeneous case under the same feedback constraint in our simulation study.

The average sum rate helps in understanding the system performance with perfect feedback. In practical feedback systems, imperfections occur
such as channel estimation error and feedback delay. These inevitable factors degrade the system performance by causing outage
\cite{piantanida09, isukapalli10}. Therefore, rather than using average sum rate as the performance metric, we employ the notion of average
goodput \cite{lau08, wu10, akoum10} to incorporate outage probability. Under the multi-cluster subband fading model, we perform analysis on the
average goodput and the average outage probability with heterogeneous partial feedback. In addition to examining the impact of
imperfect feedback on multiuser diversity \cite{ma05, kuhne08}, we also investigate how to adapt and optimize the average goodput in the presence of
these imperfections. We consider both the fixed rate and the variable rate scenarios, and utilize bounding technique and an efficient
approximation to derive near-optimal strategies.

To summarize, the contributions of this paper are threefold: a conceptual heterogeneous feedback design framework to adapt feedback amount to
the underlying channel statistics, a thorough analysis of both perfect and imperfect feedback systems under the multi-cluster subband fading
model, and the development of approximations and near-optimal approaches to adapt and optimize the system performance. The rest of the paper is
organized as follows. The motivation under the general correlated channel model and the development of system model is presented in Section
\ref{system}. Section \ref{perfect} deals with perfect feedback, and Section \ref{imperfect} examines imperfect feedback due to channel
estimation error and feedback delay. Numerical results are presented in Section \ref{numerical}. Finally, Section \ref{conclusion} concludes the
paper.

\section{System Model}\label{system}

\subsection{Motivation for Heterogeneous Partial Feedback}\label{motivation}
This part provides justification for the adaptation of subband size with one cluster of users under the general correlated channel model, and
motivates the design of heterogeneous partial feedback for the multi-cluster scenario in Section \ref{multicluster}. Consider a downlink
multiuser OFDMA system with one base station and $K$ users. One cluster of user is assumed in this part and users in this cluster are assumed to
experience the same frequency selectivity. The system consists of $N_\mathsf{c}$ subcarriers. $H_{k,n}$, the frequency domain channel transfer
function between transmitter and user $k$ at subcarrier $n$,  can be written as:
\begin{equation} \label{system:eq_1}
H_{k,n}=\sum_{l=0}^{L-1}\sigma_lF_{k,l}\exp\left(-\frac{j2\pi ln}{N_\mathsf{c}}\right),
\end{equation}
where $L$ is the number of channel taps, $\sigma_l$ for $l=0,\ldots,L-1$ represents the channel power delay profile and is normalized, i.e.,
$\sum_{l=0}^{L-1}\sigma_l^2=1$, $F_{k,l}$ denotes the discrete time channel impulse response, which is modeled as complex Gaussian distributed
random processes with zero mean and unit variance $\mathcal{CN}(0,1)$ and is i.i.d. across $k$ and $l$. Only fast fading effect is considered in
this paper, i.e., the effects of path loss and shadowing are assumed to be ideally compensated by power control\footnote[2]{This assumption has
been employed in \cite{kuhne08, leinonen09, pugh10} to simplify the scheduling policy. With the same average SNR, the opportunistic scheduling
policy is also long-term fair. When different average SNR is assumed, the proportional-fair scheduling policy \cite{viswanath02} can be
utilized.}. The received signal of user $k$ at subcarrier $n$ can be written as:
\begin{equation} \label{system:eq_2}
u_{k,n}=\sqrt{P_{\mathsf{c}}}H_{k,n}s_{k,n}+v_{k,n},
\end{equation}
where $P_{\mathsf{c}}$ is the average received power per subcarrier, $s_{k,n}$ is the transmitted symbol and $v_{k,n}$ is the additive white
noise distributed as $\mathcal{CN}(0,\sigma_{n_{\mathsf{c}}}^2)$. From (\ref{system:eq_1}), it can be shown that $H_{k,n}$ is distributed as
$\mathcal{CN}(0,1)$. The channels at different subcarriers are correlated, and the correlation coefficient between subcarriers $n_1$ and $n_2$
can be described as follows:
\begin{equation} \label{system:eq_3}
\mathrm{cov}(H_{k,n_1},H_{k,n_2})=\sum_{l=0}^{L-1}\sigma_l^2\exp\left(-\frac{j2\pi l(n_2-n_1)}{N_\mathsf{c}}\right).
\end{equation}

In general, adjacent subcarriers are highly correlated. In order to reduce feedback needs, $R_\mathsf{c}$ subcarriers are formed as one resource
block, and $\eta$ resource blocks are grouped into one subband\footnote[3]{E.g., in LTE, one resource block consists of $12$ subcarriers, and
one subband can contain $1$ to $8$ resource blocks \cite{dahlman11}.}. Thus, there are $N=\frac{N_\mathsf{c}}{R_\mathsf{c}}$ resource blocks and
$\frac{N}{\eta}$ subbands\footnote[4]{Throughout the paper, $N_\mathsf{c}$, $N$ and $\eta$ are assumed to be a radix $2$ number. A more general
treatment is possible but this will result in edge effects making for more complex notation without much insight.}. In this manner, each user
performs subband-based feedback to enable opportunistic scheduling at the transmitter. Since the channels are correlated and there is one CQI to
represent a given subband, the CQI is a function of the all the individual channels within that subband. Herein, we employ the following subband
(aggregate) average rate $S_{k,r}$ as the functional form\footnote[5]{This functional form employs the capacity formula and the resulting
effective SNR has a geometric mean interpretation. Other functional forms of the CQI exist in practical systems such as exponential effective
SNR mapping (EESM) \cite{ericsson03, song11, donthi11j} and mutual information per bit (MMIB) \cite{wan06, fan11} to map the effective SNR to
the block-error-rate (BLER) curve. The intuitions are similar: to obtain a representative CQI as a single performance measure corresponding to
the rate performance.} \cite{forney98, al96} of the CQI for user $k$ at subband $r$:
\begin{equation} \label{system:eq_4}
S_{k,r}\triangleq\frac{1}{\eta R_{\mathsf{c}}}\sum_{n=(r-1)\eta R_{\mathsf{c}}+1}^{r\eta
R_{\mathsf{c}}}\log_2\left(1+\frac{P_{\mathsf{c}}|H_{k,n}|^2}{\sigma_{n_{\mathsf{c}}}^2}\right).
\end{equation}

Each user employs the best-M partial feedback strategy and conveys back the $M$ best CQI values selected from $S_{k,r}, 1\leq r\leq
\frac{N}{\eta}$. A detailed description of the best-M strategy can be found in \cite{choi08, leinonen09, hur11}. After the base station receives
feedback, it performs opportunistic scheduling and selects the user $k$ for transmission at subband $r$ if user $k$ has the largest CQI at
subband $r$. Also, it is assumed that if no user reports CQI for a certain subband, scheduling outage happens and the transmitter does not
utilize it for transmission.

\begin{figure}[t]
\centering
    \includegraphics[width=0.55\linewidth]{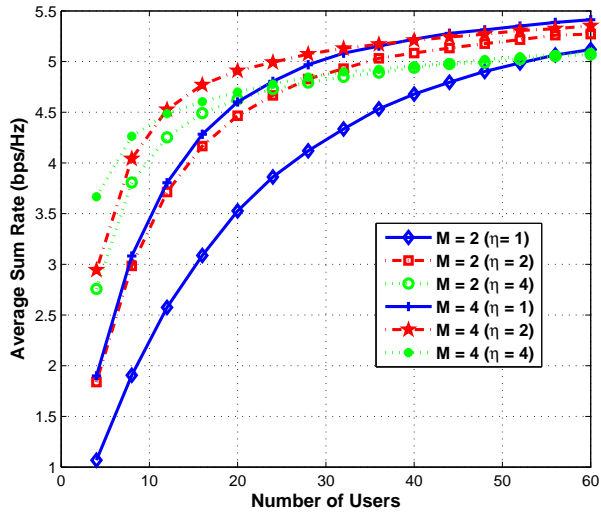}
\caption{Comparison of average sum rate for different subband sizes ($\eta=1,2,4$) and partial feedback ($M=2,4$) with respect to the number of
users. A general correlated channel model is assumed with an exponential power delay profile. ($N_{\mathsf{c}}=256$, $N=32$, $L=16$, $\delta=4$,
$\frac{P_{\mathsf{c}}}{\sigma_{n_{\mathsf{c}}}^2}=10$ dB)} \label{fig_1}
\end{figure}

\begin{figure}[t]
\centering
    \includegraphics[width=0.8\linewidth]{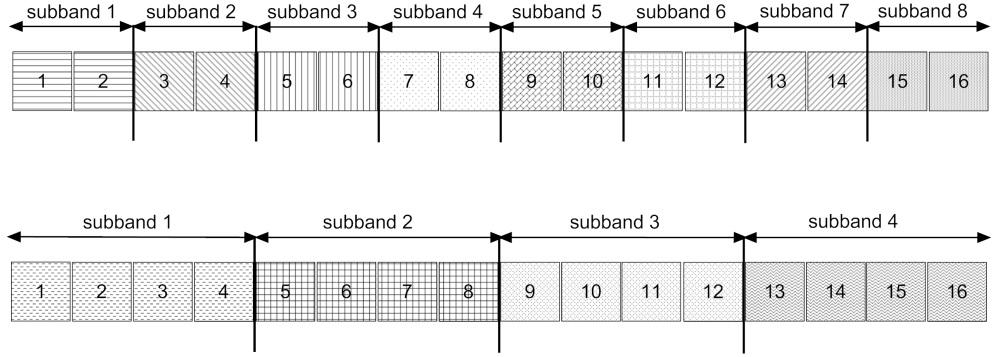}
\caption{Illustration of the multi-cluster subband fading channel model for two different clusters with $16$ resource blocks. The subband sizes
equal $2$ and $4$ for the two different clusters respectively. According to the subband fading model, the channel frequency selectivity is flat
within each subband, and independent across subbands. The subband sizes can be heterogeneous across clusters, and this leads to heterogeneous
channel frequency selectivity across clusters. The subband fading model approximates the general correlated channel model, and is useful for
statistical analysis.} \label{fig_2}
\end{figure}

Now we demonstrate the need to adapt the subband size to achieve the potential ``match" among coherence bandwidth, subband size and partial
feedback through a simulation example. The channel is modeled according to the exponential power delay profile \cite{weinfurtner02, mckay08,
eslami11}: $\sigma_l^2=\frac{1-\exp(-1/\delta)}{1-\exp(-L/\delta)}\exp\left(-\frac{l}{\delta}\right)$ for $0\leq l<L$, where the parameter
$\delta$ is related to the RMS delay spread. The simulation parameters are: $N_{\mathsf{c}}=256$, $N=32$, $L=16$, $\delta=4$,
$\frac{P_{\mathsf{c}}}{\sigma_{n_{\mathsf{c}}}^2}=10$ dB. The subband size $\eta$ can be adjusted and ranges from $1$ to $4$ resource blocks. We
consider partial feedback with $M=2$ and $M=4$. The average sum rate of the system for different subband sizes and partial feedback with respect
to the number of users is shown in Fig. \ref{fig_1}. Under the given coherence bandwidth, several observations can be made. Firstly, the curves
with $\eta=4$ has the smallest increasing rate because a larger subband size gives a poor representation of the channel. Secondly, the curve
with $\eta=1,M=2$ has the smallest average sum rate because a small amount of partial feedback is not enough to reflect the channel quality.
Thirdly, the two curves $\eta=1,M=4$ and $\eta=2,M=2$ possess similar increasing rate. This is because the underlying channel is highly
correlated within $2$ resource blocks and thus having $M$-best feedback with $\eta=2$ yields similar effect as having $2M$-best feedback with
$\eta=1$. From the above observations, $\eta=2$ matches the frequency selectivity and there would be performance loss or waste of feedback
resource when a subband size is blindly chosen. In a multi-cluster scenario where users in different clusters experience diverse coherence
bandwidth, this advocates heterogeneous subband size and heterogeneous feedback.

The general correlated channel model as well as the non-linearity of the CQI, though useful to demonstrate the need for heterogeneous feedback,
does not lend itself to tractable statistical analysis. To develop a tractable analytical framework, an approximated channel model is needed. A
widely used model is the block fading model in the frequency domain \cite{mceliece84, medard02} due to its simplicity and capability to provide
a good approximation to actual physical channels. According to the block fading model, the channel frequency selectivity is flat within each
block, and independent across blocks \cite{chen08, leinonen09, hur11}. Herein, we generalize the block fading model to the subband fading model
for the multi-cluster scenario. We assume that users possessing similar frequency selectivity are grouped into a cluster and the subband size is
perfectly matched to the coherence bandwidth for a given cluster\footnote[6]{In practical systems, since the coherence bandwidth is determined
by the channel statistics which vary on the order of tens of seconds or more, the cluster information can be learned and updated through
infrequent user feedback. Therefore, the cluster is formed dynamically but in a slow way compared to the time variation of the fast fading
effect which is on the order of milliseconds.}. According to the subband fading model, for a given cluster with a perfectly matched subband
size, the channel frequency selectivity is flat within each subband, and independent across subbands. Fig. \ref{fig_2} demonstrates the subband
fading model for two different clusters with different subband sizes under a given number of resource blocks.

\subsection{Multi-Cluster Subband Fading Model}\label{multicluster}
We now present the multi-cluster subband fading model. Consider a downlink multiuser OFDMA system with one base station and $G$ clusters of
users. The system consists of $N$ resource blocks and the total number of users equals $K$. Users in cluster $\mathcal{K}_g$ are indexed by the
set $\mathcal{K}_g=\{1,\ldots,k,\ldots,K_g\}$ for $1\leq g\leq G$, with $|\mathcal{K}_g|=K_g$ and $\sum_{g=1}^G K_g=K$. In our framework, users
in the same cluster group their resource blocks into subbands in the same manner while each cluster can potentially employ a different grouping
which enables the subband size to be heterogeneous between clusters. Denote $\eta_g$ as the subband size for cluster $\mathcal{K}_g$, and
$\eta_g \in \{2^0,2^1,\ldots,N\}$. The $\eta_g$'s are ordered such that $\eta_1<\cdots<\eta_G$. Based on the assumption for $\eta_g$, the number
of subbands in cluster $\mathcal{K}_g$ equals $\frac{N}{\eta_g}$.

Let $H_{k,r}^{(g)}$ be the frequency domain channel transfer function between transmitter and user $k$ in cluster $\mathcal{K}_g$ at subband
$r$,  where $1\leq k \leq K_g, 1\leq r \leq \frac{N}{\eta_g}$. $H_{k,r}^{(g)}$ is distributed as $\mathcal{CN}(0,1)$. According to the subband
fading model, $H_{k,r}^{(g)}$ is assumed to be independent across users and subbands in cluster $\mathcal{K}_g$. The feedback for different
clusters is at different granularity, and so to model the channel for the different clusters of users at the same basic resource block level,
some additional notation is needed. Let $\tilde{H}_{k,n}^{(g)}=H_{k,\lceil \frac{n}{\eta_g}\rceil}^{(g)}$ with $1\leq n \leq N$ denote the
resource block based channel transfer function. Then the received signals of user $k$ in cluster $\mathcal{K}_g$ at resource block $n$ can be
represented by:
\begin{equation} \label{system:eq_5}
u_{k,n}^{(g)}= \sqrt{P}\tilde{H}_{k,n}^{(g)}s_{k,n}^{(g)}+v_{k,n}^{(g)},
\end{equation}
where $P$ is the average received power per resource block, $s_{k,n}^{(g)}$ is the transmitted symbol and $v_{k,n}^{(g)}$ is additive white
noise distributed with $\mathcal{CN}(0,\sigma_n^2)$.

Let $Z_{k,r}^{(g)}\triangleq |H_{k,r}^{(g)}|^2$ denote the CQI for user $k$ in cluster $\mathcal{K}_g$ at subband $r$. In order to reduce the
feedback load, users employ the best-M strategy to feed back their CQI. In the basic best-M feedback policy, users measure CQI for each resource
block at their receiver and feed back the CQI values of the $M$ best resource blocks chosen from the total $N$ values. For each resource block,
the scheduling policy selects the user with the largest CQI among the users who fed back CQI to the transmitter for that resource block.
However, in our heterogeneous partial feedback framework, since the number of independent CQI for cluster $\mathcal{K}_g$ is $\frac{N}{\eta_g}$,
a fair and reasonable way to allocate the feedback resource is to linearly scale the feedback amount for users in cluster $\mathcal{K}_g$. To be
specific, user $k$ in $\mathcal{K}_G$ (i.e., the cluster with the largest subband size) is assumed to feed back the $M$ best CQI selected from
$\{Z_{k,r}^{(G)}\}, 1\leq r \leq \frac{N}{\eta_G}$, whereas user $k$ in $\mathcal{K}_g$ conveys the $\frac{\eta_G}{\eta_g}M$ best CQI selected
from $\{Z_{k,r}^{(g)}\}, 1\leq r \leq \frac{N}{\eta_g}$. After receiving feedback from all the clusters, for each resource block the system
schedules the user for transmission with the largest CQI. It is useful to note that the user feedback is based on the subband level, while the
base station schedules transmission at the resource block level.

\section{Perfect Feedback}\label{perfect}
In this section, the CQI are assumed to be fed back without any errors and the average sum rate is employed as the performance metric for system
evaluation. We derive a closed form expression for the average sum rate in Section \ref{spectral_efficiency} for the multi-cluster heterogeneous
feedback system. In Section \ref{ratio} we analyze the relationship between the sum rate ratio and the choice of the best-M.

\subsection{Derivation of Average Sum Rate}\label{spectral_efficiency}
According to the assumption, the CQI $Z_{k,r}^{(g)}$ is i.i.d. across subbands and users, and thus let $F_Z$ denote the CDF. Because only a
subset of the ordered CQI are fed back, from the transmitter's perspective, if it receives feedback on a certain resource block from a user, it
is likely to be any one of the CQI from the ordered subset. We now aim to find the CDF of the CQI seen at the transmitter side as a consequence
of partial feedback. Let $\tilde{Y}_{k,n}^{(g)}$ denote the reported CQI viewed at the transmitter for user $k$ in $\mathcal{K}_g$ at resource
block $n$. Also, let $Y_{k,r}^{(g)}$ represent the subband-based CQI seen at the transmitter for user $k$ in $\mathcal{K}_g$ at subband $r$,
then $\tilde{Y}_{k,n}^{(g)}=Y_{k,\lceil \frac{n}{\eta_g}\rceil}^{(g)}$. The following lemma describes the CDF of $\tilde{Y}_{k}^{(g)}$ (the
index $n$ is dropped for notational simplicity), which is denoted by $F_{\tilde{Y}_{k}^{(g)}}$.

\begin{lemma} \label{lemma_1}
The CDF of $\tilde{Y}_{k}^{(g)}$ is given by:
\begin{equation} \label{perfect:eq_1}
F_{\tilde{Y}_{k}^{(g)}}(x)=\sum_{m=0}^{\frac{\eta_G}{\eta_g}M-1}\xi_g(N,M,\boldsymbol\eta,m)(F_Z(x))^{\frac{N}{\eta_g}-m},
\end{equation}
where the vector $\boldsymbol\eta\triangleq(\eta_1,\cdots,\eta_g,\cdots,\eta_G)$ and
\begin{equation} \label{perfect:eq_2}
\xi_g(N,M,\boldsymbol\eta,m)=\sum_{i=m}^{\frac{\eta_G}{\eta_g}M-1}\frac{\frac{\eta_G}{\eta_g}M-i}{\frac{\eta_G}{\eta_g}M}{\frac{N}{\eta_g}\choose
i}{i\choose m}\left(-1\right)^{i-m}.
\end{equation}
\end{lemma}
\begin{proof}
The proof is provided in Appendix \ref{appenA}.
\end{proof}

Let $k_n^*$ demote the selected user at resource block $n$, then according to the scheduling policy:
\begin{equation} \label{perfect:eq_3}
k_n^*=\arg\max_{k\in \mathcal{U}_n}\;\{\tilde{Y}_{k,n}^{(1)},\cdots,\tilde{Y}_{k,n}^{(g)},\cdots,\tilde{Y}_{k,n}^{(G)}\},
\end{equation}
where $\mathcal{U}_n\triangleq\{\mathcal{U}_n^{(1)},\cdots,\mathcal{U}_n^{(g)},\cdots,\mathcal{U}_n^{(G)}\}$ is the set of users who convey
feedback for resource block $n$, with $|\mathcal{U}_n^{(g)}|=\tau_g$ representing the number of users belonging to $\mathcal{U}_n$ in cluster
$\mathcal{K}_g$. It can be easily seen that in the full feedback case, i.e., $M=M_F\triangleq\frac{N}{\eta_G}$, $|\mathcal{U}_n^{(g)}|=K_g$. For
the general case when $1\leq M<M_F$, the probability mass function (PMF) of $\mathcal{U}_n$ is given by:
\begin{equation} \label{perfect:eq_4}
\mathbb{P}(\mathcal{U}_n)=\left(\prod_{g=1}^G{K_g\choose \tau_g}\right)\left(\frac{\eta_G M}{N}\right) ^{\sum_{g=1}^G\tau_g}\left(1-\frac{\eta_G
M}{N}\right)^{K-\sum_{g=1}^G\tau_g}, \quad 0\leq \tau_g\leq K_g.
\end{equation}
\textit{Remark:} Only the largest subband size $\eta_G$ appears in the expression of $\mathbb{P}(\mathcal{U}_n)$ instead of the vector
$\boldsymbol\eta$. This is due to our heterogeneous partial feedback design to let users in cluster $\mathcal{K}_g$ convey back the
$\frac{\eta_G}{\eta_g}M$ best CQI out of $\frac{N}{\eta_g}$ values.

Now we turn to determine the conditional CDF of the CQI for the selected user at resource block $n$, conditioned on the set of users providing
CQI for that resource block. Since users are equiprobable to be scheduled according to the fair scheduling policy, the condition on $k_n^*$ is
not described explicitly, and so we denote the conditional CDF as $F_{X_n|\mathcal{U}_n}$, where $X_n|\mathcal{U}_n$ is the conditional CQI of
the selected user at resource block $n$. Notice from Lemma \ref{lemma_1} that $\tilde{Y}_{k}^{(g)}$ possess a different distribution for
different $g$ due to the heterogeneous feedback from different clusters. Using order statistics \cite{david03} yields $F_{X_n|\mathcal{U}_n}$
as:
\begin{equation} \label{perfect:eq_5}
F_{X_n|\mathcal{U}_n}(x)=\prod_{g=1}^G(F_{\tilde{Y}_{k}^{(g)}}(x))^{\tau_g}.
\end{equation}
Then the polynomial form of $F_{X_n|\mathcal{U}_n}$ can be obtained, which is stated in the following theorem.

\begin{theorem} \label{theorem_1}
The CDF of $F_{X_n|\mathcal{U}_n}$ is given by:
\begin{equation} \label{perfect:eq_6}
F_{X_n|\mathcal{U}_n}(x)=\sum_{m=0}^{\Phi(M,\boldsymbol\eta,\boldsymbol\tau)}\Theta_{G-1}(N,M,\boldsymbol\eta,\boldsymbol\tau,m)(F_Z(x))^{\sum_{g=1}^G\frac{N}{\eta_g}\tau_g-m},
\end{equation}
where the vector $\boldsymbol\tau\triangleq(\tau_1,\cdots,\tau_g,\cdots,\tau_G)$,
$\Phi(M,\boldsymbol\eta,\boldsymbol\tau)\triangleq\sum_{g=1}^G\tau_g\left(\frac{\eta_G}{\eta_g}M-1\right)$,
\begin{equation} \label{perfect:eq_7}
\Theta_{g}(N,M,\boldsymbol\eta,\boldsymbol\tau,m)=\left\{
\begin{array}{l} \mathop{\sum}\limits_{i=0}^m\Lambda_1(N,M,\boldsymbol\eta,\boldsymbol\tau,i)\Lambda_2(N,M,\boldsymbol\eta,\boldsymbol\tau,m-i),\quad g=1\\
\mathop{\sum}\limits_{i=0}^m\Theta_{g-1}(N,M,\boldsymbol\eta,\boldsymbol\tau,i)\Lambda_{g+1}(N,M,\boldsymbol\eta,\boldsymbol\tau,m-i),\quad 2\leq g<G\\
\end{array} \right.
\end{equation}
\begin{equation} \label{perfect:eq_8}
\Lambda_g(N,M,\boldsymbol\eta,\boldsymbol\tau,m)=\left\{
\begin{array}{l} (\xi_g(N,M,\boldsymbol\eta,0))^{\tau_g},\quad m=0\\ \frac{1}{m\xi_g(N,M,\boldsymbol\eta,0)}\sum_{\ell=1}^{\min\left(m,\frac{\eta_G}{\eta_g}
M-1\right)}((\tau_g+1)\ell-m)\\
\times\xi_g(N,M,\boldsymbol\eta,\ell)\Lambda_g(N,M,\boldsymbol\eta,\boldsymbol\tau,m-\ell),\quad 1\leq m<\tau_g(\frac{\eta_G}{\eta_g}M-1)\\
(\xi_g(N,M,\boldsymbol\eta,\frac{\eta_G}{\eta_g} M-1))^{\tau_g},\quad m=\tau_g(\frac{\eta_G}{\eta_g} M-1).\\ \end{array} \right.
\end{equation}
\end{theorem}
\begin{proof}
The proof is provided in Appendix \ref{appenA}.
\end{proof}

After obtaining the conditional CDF $F_{X_n|\mathcal{U}_n}$, let $C_P(M)$ denote the average sum rate and it can be computed using the following
procedure.
\begin{align}
C_P(M)&=\frac{1}{N}\sum_{n=1}^N\mathbb{E}[\log_2(1+X_n)]\notag\\
&\mathop{=}\limits^{(a)}\mathbb{E}_{\mathcal{U}}\left[\int_0^\infty \log_2(1+\rho
x)d(F_{X|\mathcal{U}}(x))\right]\notag\\
&\mathop{=}\limits^{(b)}\mathbb{E}_{\mathcal{U}}\left[\sum_{m=0}^{\Phi(M,\boldsymbol\eta,\boldsymbol\tau)}\Theta_{G-1}(N,M,\boldsymbol\eta,\boldsymbol\tau,m)\int_0^\infty
\log_2(1+\rho
x)d(F_Z(x))^{\sum_{g=1}^G\frac{N}{\eta_g}\tau_g-m}\right]\notag\\
\label{perfect:eq_9}&\mathop{=}\limits^{(c)}\sum_{\boldsymbol\tau\neq\mathbf{0}}\mathbb{P}(\mathcal{U})
\sum_{m=0}^{\Phi(M,\boldsymbol\eta,\boldsymbol\tau)}\Theta_{G-1}(N,M,\boldsymbol\eta,\boldsymbol\tau,m)\mathcal{I}_1\left(\rho,\sum_{g=1}^G\frac{N}{\eta_g}\tau_g-m\right),
\end{align}
where $\rho\triangleq \frac{P}{\sigma_n^2}$ and $\mathbb{P}(\mathcal{U})$ is given by (\ref{perfect:eq_4}). (a) follows from the conditional
expectation of $X_n|\mathcal{U}_n$ and the identically distributed property (let $X$ and $\mathcal{U}$ represent $X_n$ and $\mathcal{U}_n$
respectively), (b) follows from (\ref{perfect:eq_6}) in Theorem \ref{theorem_1}, (c) follows from (\ref{perfect:eq_4}), and define
$\mathcal{I}_1(a,b)\triangleq \int_0^\infty \log_2(1+ax)d(F_Z(x))^b$. $\mathcal{I}_1(a,b)$ is computed in Appendix \ref{appenA} to be:
\begin{equation}
\label{perfect:eq_10}
\mathcal{I}_1(a,b)=\frac{b}{\ln2}\sum_{\ell=0}^{b-1}\binom{b-1}{\ell}\frac{(-1)^{\ell}}{\ell+1}\exp\left(\frac{\ell+1}{a}\right)E_1\left(\frac{\ell+1}{a}\right),
\end{equation}
where $E_1(x)=\int_x^\infty \exp(-t)t^{-1}dt$ is the exponential integral function \cite{abramowitz72}.

The average sum rate for the full feedback is a special case and is given by:
\begin{equation}
\label{perfect:eq_11} C_P(M_F)=\int_0^\infty \log_2(1+\rho x)d(F_Z(x))^K=\mathcal{I}_1(\rho,K).
\end{equation}

\textit{Remark:} It is noteworthy to mention that the functional form of $C_P(M)$ in (\ref{perfect:eq_9}) consists of two main parts. The first
part, which involves $\mathbb{P}(\mathcal{U})$ and $\Theta_{G-1}(\cdot,\cdot,\cdot,\cdot,\cdot)$, accounts for the randomness of the set of
users who convey feedback as well as the scheduling policy. This part is inherent to the heterogeneous partial feedback strategy, and is
independent of the system metric for evaluation, such as the average sum rate employed in this paper. The second part
$\mathcal{I}_1(\cdot,\cdot)$ depends on statistical assumption of the underlying channel and the system metric, and it is impacted by partial
feedback as well.

\subsection{Sum Rate Ratio and Best-M}\label{ratio}
We now examine how to determine the smallest $M$ that results in almost the same performance, in terms of average sum rate, as the full feedback
case. Applying the same technique as in \cite{choi08, hur11}, define $\gamma_P$ as the spectral efficiency ratio and the problem can be
formulated as:
\begin{equation}
\label{perfect:eq_12} \mathrm{Find\; the\; minimum\;} M^{\ast},\quad s.t.\; \gamma_P=\frac{C_P(M^{\ast})}{C_P(M_F)}\geq \gamma.
\end{equation}
The above problem can be numerically solved by substituting the expressions for $C_P(M)$ and $C_P(M_F)$. In order to obtain a simpler and
tractable relationship between $M$ and $K$ given $\boldsymbol\eta$, i.e., the tradeoff between the amount of partial feedback and the number of
users given existing heterogeneity of channel statistics in frequency domain, an approximation is utilized similar to that in \cite{hur11}, by
observing that $\mathcal{I}_1(a,b)$ in (\ref{perfect:eq_10}) is slowly increasing in $b$ with fixed $a$ (This phenomenon is due to the
saturation of multiuser diversity \cite{sharif05}). Observing
$\sum_{m=0}^{\Phi(M,\boldsymbol\eta,\boldsymbol\tau)}\Theta_{G-1}(N,M,\boldsymbol\eta,\boldsymbol\tau,m)=1$ and employing the binomial theorem
yields the approximation for the spectral efficiency ratio as:
\begin{equation}
\label{perfect:eq_13} \gamma_P\simeq 1-\left(1-\frac{\eta_G M^{\ast}}{N}\right)^K.
\end{equation}
From (\ref{perfect:eq_12}) and (\ref{perfect:eq_13}), the minimum required $M^{\ast}$ can be obtained as follows:
\begin{equation} \label{perfect:eq_14}
M^{\ast}\geq \frac{N}{\eta_G}\left(1-\left(1-\gamma\right)^{\frac{1}{K}}\right).
\end{equation}
\textit{Remark:} It can be seen that $M^*$ depends on the system parameters ($N,K,\gamma$) as well on the largest subband size $\eta_G$. It is
also a consequence of our heterogeneous partial feedback assumption to let users in cluster $\mathcal{K}_g$ convey back the
$\frac{\eta_G}{\eta_g}M$ best CQI out of $\frac{N}{\eta_g}$ values. This results in the fact that obtaining feedback information from users
belonging to different clusters have almost the same statistical influence on scheduling performance.

\section{Imperfect Feedback}\label{imperfect}
After analyzing the heterogeneous partial feedback design with perfect feedback, we turn to examine the impact of feedback imperfections in this
section. We develop the imperfect feedback model due to channel estimation error and feedback delay in Section \ref{imperfect_model}, and
investigate the influence of imperfections on two different transmission strategies in Section \ref{fix_rate} and \ref{vari_rate}. Then we
propose how to optimize the system performance to adapt to the imperfections in Section \ref{optimize}.

\subsection{Imperfect Feedback Model}\label{imperfect_model}
The imperfect feedback model is built upon the subband fading model for the perfect feedback case. To differentiate from the notation for the
perfect feedback case and focus on the imperfect feedback model, the resource block index is dropped. Let $h_k$ denote the frequency domain
channel transfer function of user $k$ (users in different clusters are not temporally distinguished to avoid notational overload). Due to
channel estimation error, the user only has its estimated version $\hat{h}_k$, and the relationship between $h_k$ and $\hat{h}_k$ can be modeled
as:
\begin{equation}
\label{imperfect:eq_1} h_k=\hat{h}_k+w_k,
\end{equation}
where $w_k\sim \mathcal{CN}(0,\sigma_{w_k}^2)$ is the channel estimation error. The channel of each resource block is assumed to be estimated
independently, which yields the channel estimation errors $w_k$ i.i.d. across users and resource blocks, i.e., $w_k\sim
\mathcal{CN}(0,\sigma_{w}^2)$. It is clear that the base station makes decision on scheduling and adaptive transmission depending on CQI, a
function of $\hat{h}_k$. Thus this information can be outdated due to delay between the instant CQI is measured and the actual instant of use
for data transmission to the selected user. Let $\tilde{h}_k$ be the actual channel transfer function and we employ a first-order
Gaussian-Markov model \cite{kuhne08, isukapalli10, akoum10} to describe the time evolution and to capture the relationship with the delayed
version as follows:
\begin{equation}
\label{imperfect:eq_2} \tilde{h}_k =
\alpha_k(\hat{h}_k+w_k)+\sqrt{1-\alpha_k^2}\varepsilon_k,
\end{equation}
where $\varepsilon_k$ accounts for the innovation noise and is distributed as $\mathcal{CN}(0,1)$. The delay time between $\tilde{h}_k$ and
$\hat{h}_k$ is not explicitly written for notational simplicity, and $\alpha_k\in [0,1]$ is used to model the correlation coefficient. Since the
feedback delay is mainly caused by the periodic feedback interval and processing complexity \cite{akoum10}, the innovation noise $\varepsilon_k$
are i.i.d. across users and a common $\alpha$ is assumed. Moreover, $w_k$ and $\varepsilon_k$ are assumed independent. Therefore, for notational
simplicity, the user index $k$ in the aforementioned parameters is dropped and $\hat{Z}\triangleq|\hat{h}|^2$ is denoted as CQI.

Let $\tilde{\chi},\chi$ and $\hat{\chi}$ represent: the actual CQI of the selected user for transmission, its outdated version, and its outdated
estimate respectively ($\hat{\chi}$ corresponds to $X$ for the perfect feedback case in Section \ref{spectral_efficiency}). Notice that the PDF
of the outdated estimate $\hat{\chi}$ depends on the heterogeneous feedback design and the scheduling strategy, whereas the conditional PDF of
$\tilde{\chi}|\hat{\chi}$ only depends on $\alpha$ and $\sigma_w^2$. Employing the same method in \cite{ma05, kuhne08}, the conditional PDF is
obtained as follows:
\begin{equation}
\label{imperfect:eq_3}
f_{\tilde{\chi}|\hat{\chi}}(x|\hat{\chi})=\frac{\alpha_w^2}{2}\exp\left(-\frac{\alpha_w^2x+\alpha_w^2\alpha^2\hat{\chi}}{2}\right)I_0(\alpha_w^2\alpha\sqrt{\hat{\chi}x}),
\end{equation}
where $\alpha_w=\sqrt{\frac{2}{\alpha^2\sigma_w^2+1-\alpha^2}}$, and $I_0(\cdot)$ is the zeroth-order modified Bessel function of the first kind
\cite{abramowitz72}.

Since the feedback is imperfect, there are two types of issues that arise. The first is the choice of the incorrect user to serve. However,
because of the i.i.d nature of the errors this does not compromise the fairness and also does not complicate the determination of the CDF. The
second problem is that of outage because the rate adaptation is made by the base station based on the erroneous CQI. Because of the error in the
CQI, the rate chosen may exceed the rate that the channel can support and so the base station has to take steps to mitigate this effect of
outage. A conservative strategy will result in less outage but under utilization of the channel while an aggressive strategy will result in good
utilization of the channel but only for a small fraction of the time. We now present two transmission strategies to address the outage issue.

\subsection{Fix Rate Strategy}\label{fix_rate}
In the fix rate conservative scenario, a system parameter $\beta_0$ is chosen for rate adaptation, and outage results under the following condition:
\begin{equation}
\label{imperfect:eq_4} \mathrm{Declare\; outage\; if:}\quad \{\tilde{\chi}\leq\beta_0|\hat{\chi}\}.
\end{equation}
The system average goodput is defined as the total average bps/Hz successfully transmitted \cite{lau08}. We derive the average goodput and
average outage probability for a given choice of system parameter $\beta_0$ in the following procedure.

Firstly the conditional outage probability is expressed as:
\begin{equation}
\label{imperfect:eq_5} \mathbb{P}(\tilde{\chi}\leq\beta_0|\hat{\chi})=1-\mathcal{Q}_1(\alpha_w\alpha\sqrt{\hat{\chi}},\alpha_w\sqrt{\beta_0}),
\end{equation}
where $\mathcal{Q}_1(a,b)=\int_b^{\infty}t\exp(-\frac{t^2+a^2}{2})I_0(at)dt$ is the first-order Marcum-Q function \cite{nuttall72}. Denote
$R_0(\beta_0,M)$ as the average goodput for the heterogeneous partial feedback system, which is written according to definition:
\begin{equation}
\label{imperfect:eq_6}
R_0(\beta_0,M)=\mathbb{E}_{\mathcal{U}}\left[\mathbb{E}_{\hat{\chi}|\mathcal{U}}\left[\mathbb{P}(\tilde{\chi}\geq\beta_0|\hat{\chi})\log_2(1+\rho\beta_0)\right]\right].
\end{equation}
Then, from (\ref{perfect:eq_4}) and (\ref{perfect:eq_9}), $R_0(\beta_0,M)$ can be computed as:
\begin{align}
&R_0(\beta_0,M)\notag\\
&=\mathbb{E}_{\mathcal{U}}\left[\sum_{m=0}^{\Phi(M,\boldsymbol\eta,\boldsymbol\tau)}\Theta_{G-1}(N,M,\boldsymbol\eta,\boldsymbol\tau,m)\int_0^\infty
\mathcal{Q}_1(\alpha_w\alpha\sqrt{x},\alpha_w\sqrt{\beta_0})\log_2(1+\rho\beta_0)d(F_{\hat{Z}}(x))^{\sum_{g=1}^G\frac{N}{\eta_g}\tau_g-m}\right]\notag\\
\label{imperfect:eq_7}&=\sum_{\boldsymbol\tau\neq\mathbf{0}}\mathbb{P}(\mathcal{U})\sum_{m=0}^{\Phi(M,\boldsymbol\eta,\boldsymbol\tau)}\Theta_{G-1}(N,M,\boldsymbol\eta,\boldsymbol\tau,m)\log_2(1+\rho\beta_0)\mathcal{I}_2\left(\beta_0,\sum_{g=1}^G\frac{N}{\eta_g}\tau_g-m\right),
\end{align}
where $\mathcal{I}_2(a,b)\triangleq\int_0^\infty \mathcal{Q}_1(\alpha_w\alpha\sqrt{x},\alpha_w\sqrt{a})d(F_{\hat{Z}}(x))^b$.
$\mathcal{I}_2(a,b)$ is computed in Appendix \ref{appenB} to be:
\begin{equation}
\label{imperfect:eq_8}
\mathcal{I}_2(a,b)=\frac{2b}{(1-\sigma_w^2)\ln2}\sum_{\ell=0}^{b-1}\binom{b-1}{\ell}(-1)^{\ell}\frac{1}{\zeta_{\ell}}\left(\exp(-\frac{\vartheta^2}{2})+\exp(-\frac{\zeta_{\ell}\vartheta^2}{2(\varpi^2+\zeta_{\ell})})(1-\exp(-\frac{\varpi^2\vartheta^2}{2(\varpi^2+\zeta_{\ell})}))\right),
\end{equation}
where $\varpi=\alpha_w\alpha$, $\vartheta=\alpha_w\sqrt{a}$,
$\zeta_{\ell}=\frac{2(\ell+1)}{1-\sigma_w^2}$.

The average outage probability $P_0(\beta_0,M)$ for the heterogeneous partial feedback design can be directly calculated from definition and
(\ref{imperfect:eq_7}) as follows:
\begin{align}
P_0(\beta_0,M)&=\mathbb{E}_{\mathcal{U}}\left[\mathbb{E}_{\hat{\chi}|\mathcal{U}}\left[\mathbb{P}(\tilde{\chi}\leq\beta_0|\hat{\chi})\right]\right]\notag\\
\label{imperfect:eq_9}&=\sum_{\boldsymbol\tau\neq\mathbf{0}}\mathbb{P}(\mathcal{U})\sum_{m=0}^{\Phi(M,\boldsymbol\eta,\boldsymbol\tau)}\Theta_{G-1}(N,M,\boldsymbol\eta,\boldsymbol\tau,m)\left(1-\mathcal{I}_2\left(\beta_0,\sum_{g=1}^G\frac{N}{\eta_g}\tau_g-m\right)\right).
\end{align}

The average goodput and average outage probability for the full feedback scenario is a special case and is given by:
\begin{align}
R_0(\beta_0,M_F)&=\log_2(1+\rho\beta_0)\mathcal{I}_2(\beta_0,K),\notag\\
\label{imperfect:eq_10}P_0(\beta_0,M_F)&=1-\mathcal{I}_2(\beta_0,K).
\end{align}

\subsection{Variable Rate Strategy}\label{vari_rate}
Instead of choosing a conservative system parameter to account for the fix rate scenario as in the previous subsection, we consider an approach
we refer to as the variable rate strategy. In the variable rate scenario, a system parameter $\beta_1$ is chosen and outage results under the
following condition:
\begin{equation}
\label{imperfect:eq_11} \mathrm{Declare\; outage\; if:}\quad \{\tilde{\chi}\leq\beta_1\hat{\chi}|\hat{\chi}\},
\end{equation}
where $\beta_1$ can be regarded as the backoff factor. The system average goodput and average outage probability can be derived utilizing the
following procedure.

Now under the variable rate scenario, the conditional outage probability is expressed as:
\begin{equation}
\label{imperfect:eq_12}
\mathbb{P}(\tilde{\chi}\leq\beta_1\hat{\chi}|\hat{\chi})=1-\mathcal{Q}_1(\alpha_w\alpha\sqrt{\hat{\chi}},\alpha_w\sqrt{\beta_1\hat{\chi}}).
\end{equation}
Using the same method as (\ref{imperfect:eq_6}) and (\ref{imperfect:eq_7}), let $R_1(\beta_1,M)$ denote the average goodput for the variable
rate scenario whose expression can be written as follows:
\begin{align}
\label{imperfect:eq_13}
R_1(\beta_1,M)&=\mathbb{E}_{\mathcal{U}}\left[\mathbb{E}_{\hat{\chi}|\mathcal{U}}\left[\mathbb{P}(\tilde{\chi}\geq\beta_1\hat{\chi}|\hat{\chi})\log_2(1+\rho\beta_1\hat{\chi})\right]\right]\notag\\
&=\sum_{\boldsymbol\tau\neq\mathbf{0}}\mathbb{P}(\mathcal{U})\sum_{m=0}^{\Phi(M,\boldsymbol\eta,\boldsymbol\tau)}\Theta_{G-1}(N,M,\boldsymbol\eta,\boldsymbol\tau,m)\mathcal{I}_3\left(\beta_1,\sum_{g=1}^G\frac{N}{\eta_g}\tau_g-m\right),
\end{align}
where $\mathcal{I}_3(a,b)\triangleq\int_0^\infty \mathcal{Q}_1(\alpha_w\alpha\sqrt{x},\alpha_w\sqrt{ax})\log_2(1+\rho ax)d(F_{\hat{Z}}(x))^b$.

For the full feedback case, the average goodput is given by:
\begin{equation}
\label{imperfect:eq_14} R_1(\beta_1,M_F)=\mathcal{I}_3(\beta_1,K).
\end{equation}

Note that unlike $\mathcal{I}_2(a,b)$, $\mathcal{I}_3(a,b)$ can not be written in closed form. Therefore, bounding technique and suitable
approximation are attractive to find closed form alternatives for $\mathcal{I}_3(a,b)$. The following proposition presents a valid closed form
upper bound for $\mathcal{I}_3(a,b)$ in the low $\mathsf{SNR}$ regime.
\begin{proposition}
\label{proposition_1} In the low $\mathsf{SNR}$ regime, $\mathcal{I}_3(a,b)$ can be efficiently upper bounded by:
\begin{align}
\mathcal{I}_3^{\mathrm{UB}}(a,b)=&\frac{4\rho ab}{(1-\sigma_w^2)\ln2}\sum_{\ell=0}^{b-1}(-1)^{\ell}\frac{1}{\zeta_{\ell}^2}\Bigg(1+\frac{\vartheta^2}{\varphi_{\ell}}\Bigg(\frac{\varpi^2}{\varphi_{\ell}}{}_2F_1\left(1,\frac{3}{2};2;\frac{4\varpi^2\vartheta^2}{\varphi_{\ell}^2}\right)-{}_2F_1\left(\frac{1}{2},1;1;\frac{4\varpi^2\vartheta^2}{\varphi_{\ell}^2}\right)\notag\\
\label{imperfect:eq_15}&+\frac{2\zeta_{\ell}}{\varphi_{\ell}}\Bigg(\frac{\varpi^2}{\varphi_{\ell}}{}_2F_1\left(\frac{3}{2},2;2;\frac{4\varpi^2\vartheta^2}{\varphi_{\ell}^2}\right)-\frac{1}{2}{}_2F_1\left(1,\frac{3}{2};1;\frac{4\varpi^2\vartheta^2}{\varphi_{\ell}^2}\right)\Bigg)\Bigg)\Bigg),
\end{align}
where $\varpi=\alpha_w\alpha$, $\vartheta=\alpha_w\sqrt{a}$, $\zeta_{\ell}=\frac{2(\ell+1)}{1-\sigma_w^2}$,
$\varphi_{\ell}=\varpi^2+\vartheta^2+\zeta_{\ell}$, and ${}_2F_1(\cdot,\cdot;\cdot;\cdot)$ is the Gaussian hypergeometric function
\cite{abramowitz72}.
\end{proposition}
\begin{proof}
The proof is provided in Appendix \ref{appenB}.
\end{proof}
$\mathcal{I}_3^{\mathrm{UB}}(a,b)$ is valid and tight especially for the low $\mathsf{SNR}$ regime. In order to track $\mathcal{I}_3(a,b)$ over
the whole $\mathsf{SNR}$ regimes, we propose the following approximation method by leveraging Jensen's inequality \cite{boyd04}. Recall the
definition of $\mathcal{I}_3(a,b)=\mathbb{E}[\mathcal{Q}_1(\alpha_w\alpha\sqrt{\check{\chi}},\alpha_w\sqrt{a\check{\chi}})\log_2(1+\rho
a\check{\chi})]$, where the random variable $\check{\chi}$ is defined to have CDF $(F_{\hat{Z}}(x))^b$. Firstly, $\mathbb{E}[\check{\chi}]$ can
be computed and is given by:
\begin{align}
\mathbb{E}[\check{\chi}]&=\int_0^\infty
x\frac{b}{1-\sigma_w^2}\sum_{\ell=0}^{b-1}\binom{b-1}{\ell}(-1)^{\ell}\exp\left({-\frac{(\ell+1)x}{1-\sigma_w^2}}\right)dx\notag\\
\label{imperfect:eq_16}&=\frac{b}{1-\sigma_w^2}\sum_{\ell=0}^{b-1}\binom{b-1}{\ell}(-1)^{\ell}\left(\frac{1-\sigma_w^2}{\ell+1}\right)^2.
\end{align}
Then plugging (\ref{imperfect:eq_16}) into $\mathcal{Q}_1(\alpha_w\alpha\sqrt{x},\alpha_w\sqrt{ax})\log_2(1+\rho ax)$ yields:
\begin{equation}
\label{imperfect:eq_17}
\mathcal{I}_3^{A}(a,b)=\mathcal{Q}_1\left(\alpha_w\alpha\sqrt{\mathbb{E}[\check{\chi}]},\alpha_w\sqrt{a\mathbb{E}[\check{\chi}]}\right)\log_2(1+\rho
a\mathbb{E}[\check{\chi}]).
\end{equation}

\begin{figure}[t]
\centering
    \includegraphics[width=0.68\linewidth]{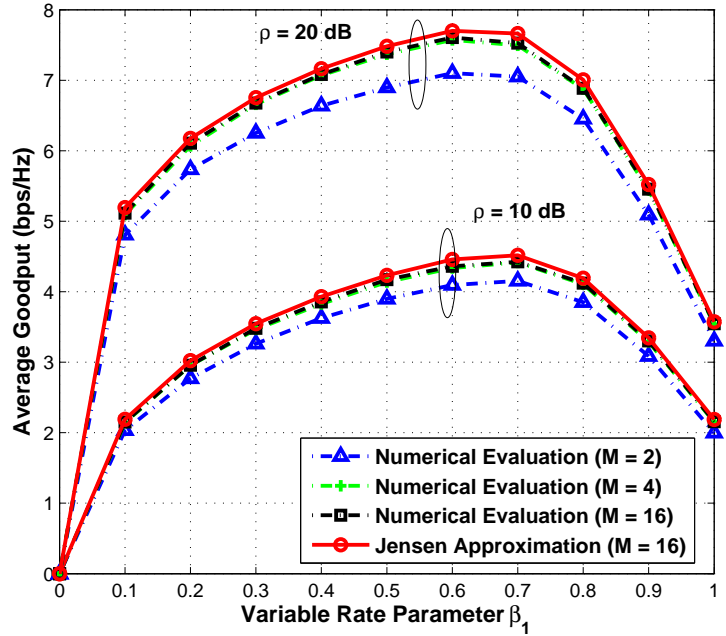}
\caption{Calculating the average goodput from numerical evaluation ($M=2,4,16$) and Jensen approximation ($M=16$) for the variable rate scenario
under different $\rho$. ($N=64$, $\boldsymbol\eta=(1,4)$, $K_1=K_2=K/2=10$, $\alpha=0.98$, $\sigma_w^2=0.01$, $\rho=10$ dB, and $20$ dB)}
\label{fig_3}
\end{figure}

Note that $\mathcal{I}_3^{A}(a,b)$ would serve as an upper bound from Jensen's inequality if the function of interest
$\mathcal{Q}_1(\alpha_w\alpha\sqrt{x},\alpha_w\sqrt{ax})\log_2(1+\rho ax)$ was concave in $x$. Properties of this function such as monotonicity
and concavity are of interest and lead to rigorous arguments in support of this bound. If outage does not occur, extensive analysis can be
carried out due to the well known properties of the $\log(\cdot)$ function. However, the concavity (or log-concavity) of
$\mathcal{Q}_1(\alpha_w\alpha\sqrt{x},\alpha_w\sqrt{\beta_1x})$ in $x$ (notice that $x$ appears in both entries of $\mathcal{Q}_1(\cdot,\cdot)$)
still remains an important open problem \cite{yu11}. Our numerical evidence suggests that
$\mathcal{Q}_1(\alpha_w\alpha\sqrt{x},\alpha_w\sqrt{\beta_1x})\log_2(1+\rho \beta_1x)$ is concave and monotonically increasing in $x$ for
practical choices of $\beta_1$.  For any given $\beta_1$ preserving the aforementioned property, Jensen's inequality yields an upper bound,
whose tightness is of interest and discussed in the following proposition. The word practical is used to exclude the situation when $\beta_1$
approaches its maximum $1$ which in turn enables $\mathcal{Q}_1(\cdot,\cdot)$ to dominate the goodput to incur extreme outage. This makes
intuitive sense according to the definition of average goodput.
\begin{proposition}
\label{proposition_2} Let $\{\check{\chi}_b\}$ be the family of positive i.i.d. random variables. If
$\mathcal{Q}_1(\alpha_w\alpha\sqrt{x},\alpha_w\sqrt{\beta_1x})\log_2(1+\rho \beta_1x)$ is concave and monotonically increasing in $x$ for any
given $\beta_1$, then the Jensen bound is asymptotically tight, i.e., as $b\rightarrow\infty$,
\begin{equation}
\label{imperfect:eq_18} \frac{\mathbb{E}[\mathcal{Q}_1(\alpha_w\alpha\sqrt{\check{\chi}_b},\alpha_w\sqrt{\beta_1\check{\chi}_b})\log_2(1+\rho
\beta_1\check{\chi}_b)]}{\mathcal{Q}_1(\alpha_w\alpha\sqrt{\mathbb{E}[\check{\chi}_b]},\alpha_w\sqrt{\beta_1\mathbb{E}[\check{\chi}_b]})\log_2(1+\rho
\beta_1\mathbb{E}[\check{\chi}_b])}\rightarrow 1.
\end{equation}
\begin{proof}
The proof is provided in Appendix \ref{appenB}.
\end{proof}
\end{proposition}
Nonetheless, when the aforementioned property is not preserved (e.g., $\beta_1$ approaches $1$), Jensen's inequality does not hold but the
expression has been experimentally found to be a good approximation and so can still be used. Therefore, (\ref{imperfect:eq_17}) is denoted as
Jensen approximation. We conduct a numerical study and demonstrate the tightness of Jensen approximation in Fig. \ref{fig_3}. It is observed
that the approximation method is very tight for moderate (even small) number of users and for all values of $\beta_1\in[0,1]$, which shows its
potential in accurately tracking the performance of average goodput.

Now we calculate the average outage probability. Since it does not involve the $\log(\cdot)$ function, it can be computed into closed form as
follows:
\begin{align}
P_1(\beta_1,M)&=\mathbb{E}_{\mathcal{U}}\left[\mathbb{E}_{\hat{\chi}|\mathcal{U}}\left[\mathbb{P}(\tilde{\chi}\leq\beta_1\hat{\chi}|\hat{\chi})\right]\right]\notag\\
\label{imperfect:eq_19}&=\sum_{\boldsymbol\tau\neq\mathbf{0}}\mathbb{P}(\mathcal{U})\sum_{m=0}^{\Phi(M,\boldsymbol\eta,\boldsymbol\tau)}\Theta_{G-1}(N,M,\boldsymbol\eta,\boldsymbol\tau,m)\left(1-\mathcal{I}_4\left(\beta_1,\sum_{g=1}^G\frac{N}{\eta_g}\tau_g-m\right)\right),
\end{align}
where \begin{align}
\mathcal{I}_4(a,b)&\triangleq\int_0^\infty
\mathcal{Q}_1(\alpha_w\alpha\sqrt{x},\alpha_w\sqrt{ax})d(F_{\hat{Z}}(x))^b\notag\\
&\mathop{=}\limits^{(a)}\frac{2b}{(1-\sigma_w^2)}\sum_{\ell=0}^{b-1}\binom{b-1}{\ell}(-1)^{\ell}\int_0^\infty\mathcal{Q}_1(\alpha_w\alpha x,\alpha_w\sqrt{a}x)\exp\left(-\frac{(\ell+1)x^2}{1-\sigma_w^2}\right)xdx\notag\\
\label{imperfect:eq_20}&\mathop{=}\limits^{(b)}\frac{b}{(1-\sigma_w^2)}\sum_{\ell=0}^{b-1}\binom{b-1}{\ell}(-1)^{\ell}\frac{1}{\zeta_{\ell}}\left(1+\frac{\psi_{\ell}}{\varsigma_{\ell}}\right),
\end{align}
$\varpi=\alpha_w\alpha$, $\vartheta=\alpha_w\sqrt{a}$, $\zeta_{\ell}=\frac{2(\ell+1)}{1-\sigma_w^2}$,
$\varphi_{\ell}=\varpi^2+\vartheta^2+\zeta_{\ell}$, $\psi_{\ell}=\varpi^2-\vartheta^2+\zeta_{\ell}$,
$\varsigma_{\ell}=\sqrt{\varphi_{\ell}^2-4\varpi^2\vartheta^2}$. (a) follows from change of variables; (b) follows from applying
\cite[B.48]{simon02}.

In the case of full feedback, the average outage probability $P_1(\beta_1,M_F)$ becomes:
\begin{equation}
\label{imperfect:eq_21} P_1(\beta_1,M_F)=1-\mathcal{I}_4(\beta_1,K).
\end{equation}

\subsection{Optimization and Adaptation to Imperfections}\label{optimize}
We have obtained the relationship between the system parameter ($\beta_0$ or $\beta_1$) and the system average goodput, and we now aim to
maximize the average goodput by adapting the system parameters.

Consider the optimization of $R_1(\beta_1,M)$ to obtain the optimal backoff factor $\beta_1^{\ast}$. It is observed from (\ref{imperfect:eq_13})
that directly optimizing $R_1(\beta_1,M)$ is tedious, and a near-optimal method is now proposed to obtain $\beta_1^{\ast}$. This method is
inspired by the results in Section \ref{ratio}, which show that the minimum required $M^{\ast}$ can be chosen to achieve almost the same
performance as a system with full feedback. Thus an optimal $\beta_1^{\ast}$ for the full feedback scenario can be optimized first, and then
$M^{\ast}$ is obtained to ``match" the system performance. Looking again at Fig. \ref{fig_3} with emphasis on different number $M$ of partial
feedback, as $M$ gets larger, the optimal $\beta_1$ converges to the full feedback case. In this example, $M^{\ast}=4$ is adequate to match the
system performance. It is noteworthy to mention that this adaptation philosophy can be applied to partial feedback systems wherein system
parameters are optimized according to full feedback assumption first and minimum required partial feedback is chosen subsequently.

Note that a closed form approximation has been obtained to track $R_1(\beta_1,M_F)$ in Section \ref{vari_rate}, which is denoted as
$R_1^A(\beta_1,M_F)\triangleq\mathcal{I}_3^{\mathrm{A}}(\beta_1,K)$. The following proposition demonstrates the optimal property of $\beta_1$
when optimizing $R_1^A(\beta_1,M_F)$.
\begin{proposition}
\label{proposition_3} There exists a unique global optimal $\beta_1$ that maximizes $R_1^A(\beta_1,M_F)$.
\end{proposition}
\begin{proof}
The proof is provided in Appendix \ref{appenB}.
\end{proof}
From the above analysis, the optimization strategy can be described as:
\begin{equation}
\label{imperfect:eq_22} \beta_1^{\ast}=\arg\mathop{\max}\limits_{0\leq\beta_1\leq1} R_1^A(\beta_1,M_F)\simeq
\arg\mathop{\max}\limits_{0\leq\beta_1\leq1} R_1(\beta_1,M_F).
\end{equation}
Since it is proved in Proposition \ref{proposition_3} that $R_1^A(\beta_1,M_F)$ is quasiconcave \cite{boyd04} in $\beta_1$, numerical approach
such as Newton-Raphson method can be applied to obtain $\beta_1^{\ast}$. As discussed before, once $\beta_1^*$ is found, the minimum required
$M^*$ can be obtained by solving (\ref{perfect:eq_12}) or relying on (\ref{perfect:eq_14}).

The same strategy can be carried over to the optimization of $\beta_0$, which is presented as follows:
\begin{equation}
\label{imperfect:eq_23} \beta_0^{\ast}=\arg\mathop{\max}\limits_{\beta_0} R_0(\beta_0,M_F).
\end{equation}
The impact of imperfections on system parameter adaptation, and the comparison between the fixed rate and variable rate strategies will be
examined through simulations in Section \ref{numerical_imperfect}.

\section{Numerical Results}\label{numerical}
In this section, we conduct a numerical study to verify the results developed and to draw some insight.

\begin{figure}[t]
\centering
\begin{tabular}{cc}
    \includegraphics[width=0.5\linewidth]{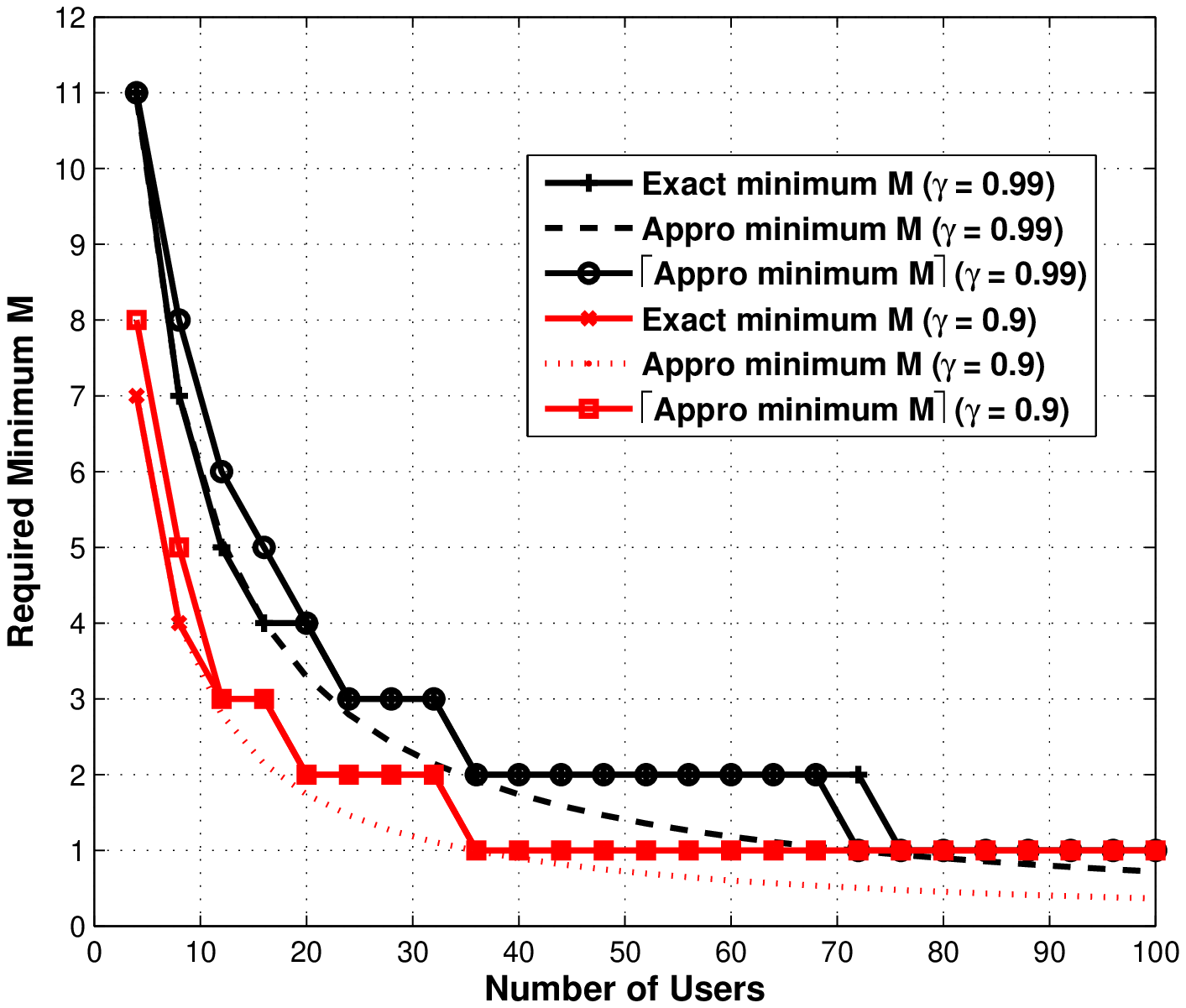}&
    \includegraphics[width=0.5\linewidth]{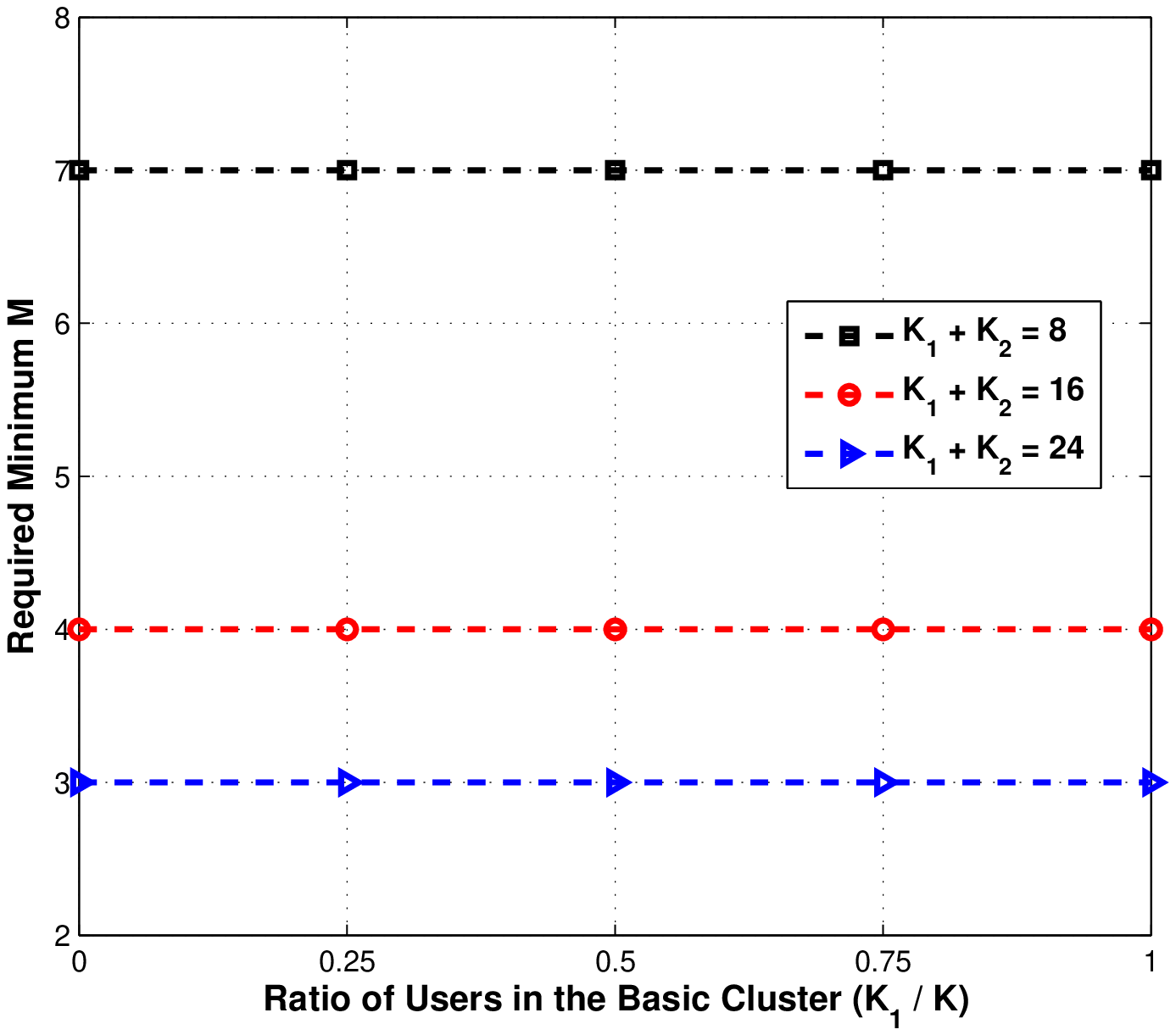}\\
    \scriptsize{(a)}&
    \scriptsize{(b)}\\
\end{tabular}
\caption{The required minimum $M$ for heterogeneous perfect feedback design: (a) Comparison of the required minimum $M$ between numerically
solving (\ref{perfect:eq_12}) and using approximation (\ref{perfect:eq_14}) under different $\gamma$ with respect to the number of users;
($N=64$, $\boldsymbol\eta=(1,4)$, $K_1=K_2=K/2$, $\rho=10$ dB) (b) Computing the required minimum $M$ with respect to different number of users
when varying the ratio of the number of users in cluster $\mathcal{K}_1$. ($N=64$, $\boldsymbol\eta=(1,4)$, $\rho=10$ dB, $\gamma=0.99$)}
\label{fig_4}
\end{figure}

\subsection{Perfect Feedback Scenario}\label{numerical_perfect}
The number of resource blocks $N$ is assumed to be $64$ for simulations throughout this section. We first consider a $2$-cluster system. Fig.
\ref{fig_4} (a) plots the minimum required $M$ obtained by directly solving (\ref{perfect:eq_12}) and alternatively by the approximation
(\ref{perfect:eq_14}) for two thresholds: $\gamma=0.99$ and $0.9$. Note that the result from (\ref{perfect:eq_14}) is rounded with the ceiling
function since the required $M$ is an integer. The other simulation parameters are $\boldsymbol\eta=(1,4)$ (i.e., $M_F=16$), and $\rho=10$ dB.
It is observed that the results from the approximate expression matches quite well with the exact computation. The question of whether the
required $M$ is sensitive to the partition of users in the system is examined in Fig. \ref{fig_4} (b) wherein the ratio of the number of users
in cluster $\mathcal{K}_1$ is changed and the minimum required $M$ is depicted for different total number of users with threshold $\gamma=0.99$.
Interestingly, the result turns out to be ``uniform". As discussed in Section \ref{perfect}, it is due to the heterogenous feedback design
assumption to let users in cluster $\mathcal{K}_1$ consume $\frac{\eta_G}{\eta_1} M$ ($4M$ in this simulation) feedback which results in the
fact that obtaining feedback information from users belonging to different clusters have almost the same influence on scheduling performance.
Therefore, the representative simulation setup $K_g=K/G$ can be employed when the system performance metric is investigated with respect to the
total number of users.

\begin{figure}[t]
\centering
    \includegraphics[width=0.7\linewidth]{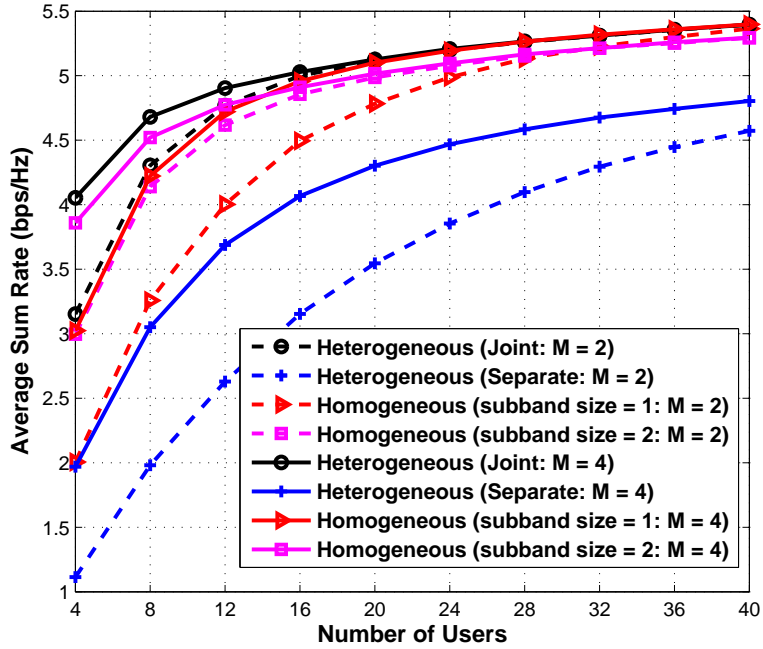}
\caption{Comparison of the average sum rate for a $4$-cluster system under different feedback strategies with respect to the number of users
($N=64$, $\boldsymbol\eta=(1,2,4,8)$, $K_1=K_2=K_3=K_4=K/4$, $M=2,4$, $\rho=10$ dB)} \label{fig_5}
\end{figure}

We now consider a $4$-cluster system with subband size vector $\boldsymbol\eta=(1,2,4,8)$ (i.e., $M_F=8$). Fig. \ref{fig_5} demonstrates the
benefit of using heterogeneous feedback design. One of the competing strategies is also heterogeneous, but treats users from each cluster
separately. In particular, the system firstly clusters the users based on their channel statistics, and then serves the clusters one by one
requiring feedback only from the served cluster of users. In this way, the feedback amount is varying over time depending on the partition of
users. This strategy is denoted as \emph{separate} heterogeneous feedback compared to our \emph{joint} heterogeneous feedback design. The other
competing strategies are homogeneous without taking advantage of the channel statistics of different users. To maintain at least the same
feedback amount for fair comparison, each user in the homogeneous case is assumed to feed back
$\lceil\sum_{g=1}^G\frac{\eta_G}{\eta_g}\frac{M}{G}\rceil$ CQI values. Two subband sizes are assumed for the homogeneous feedback. It is clear
that for the homogeneous case, users in cluster $\mathcal{K}_1$ have more independent feedback while users in cluster $\mathcal{K}_4$ suffer
from redundant feedback. The average sum rate for two different values of $M$ are shown in Fig. \ref{fig_5}. The separate heterogenous feedback
is observed to have the worst performance from a sum rate perspective because it does not fully exploit multiuser diversity, but it consumes the
least feedback. Our joint heterogenous feedback design is shown to perform much better than the two homogeneous strategies for the $4$-cluster
system. It is due to the fact that by considering the existing heterogeneity among users, the proposed heterogeneous design can make the best
use of the degrees of freedom in the frequency domain in order to enhance the system performance as well as reduce feedback needs.

\begin{figure}[t]
\centering
\begin{tabular}{cc}
    \includegraphics[width=0.5\linewidth]{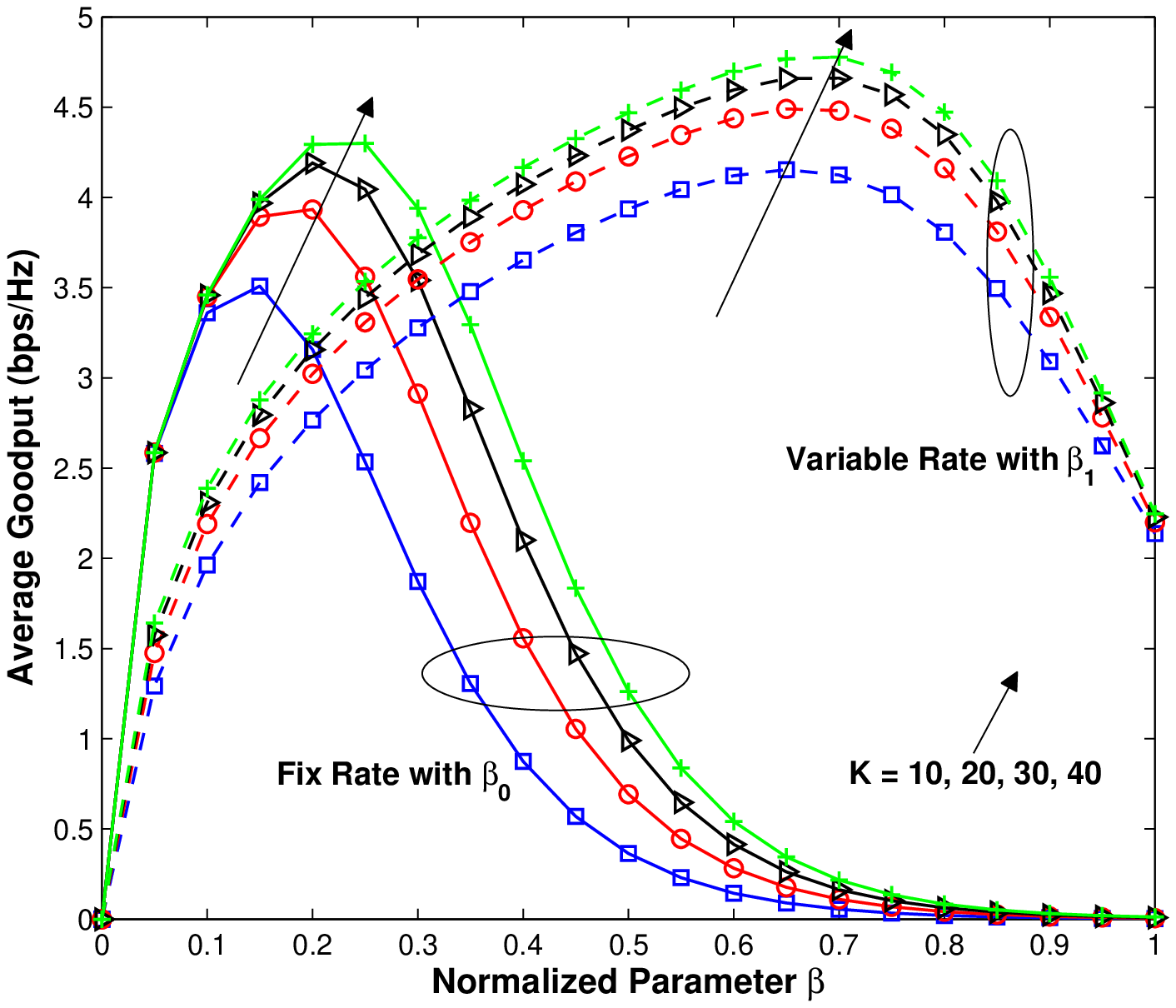}&
    \includegraphics[width=0.5\linewidth]{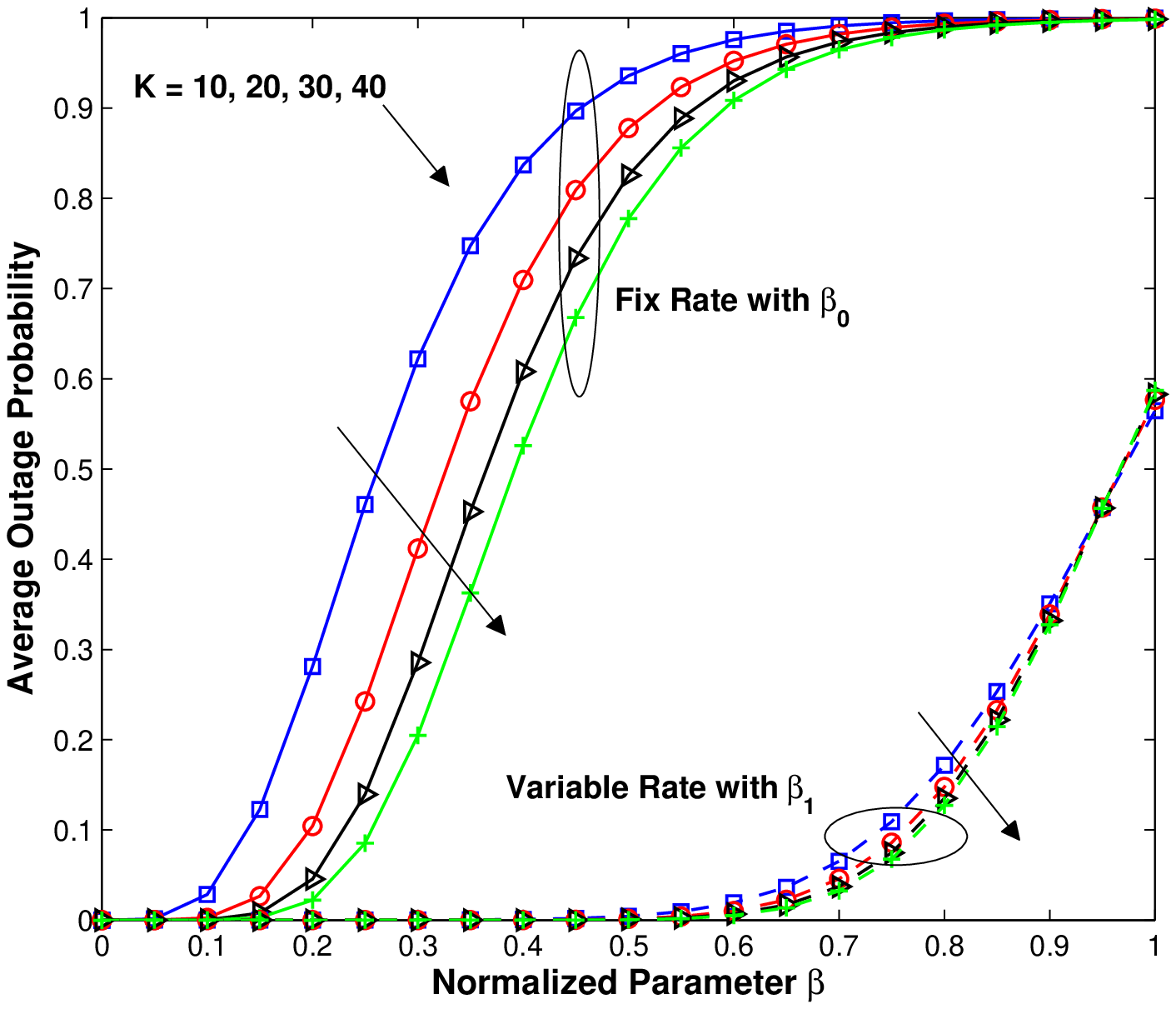}\\
    \scriptsize{(a)}&
    \scriptsize{(b)}\\
\end{tabular}
\caption{Comparison of fixed rate and variable rate strategies under normalized parameter $\beta$ ($\beta=\beta_1=\beta_0/10$) for different
number of users $K$ ($N=64$, $\alpha=0.98$, $\sigma_w^2=0.01$, $\rho=10$ dB): (a) Comparison of average goodput; (b) Comparison of average
outage probability.} \label{fig_6}
\end{figure}

\begin{figure}[t]
\centering
\begin{tabular}{cc}
    \includegraphics[width=0.5\linewidth]{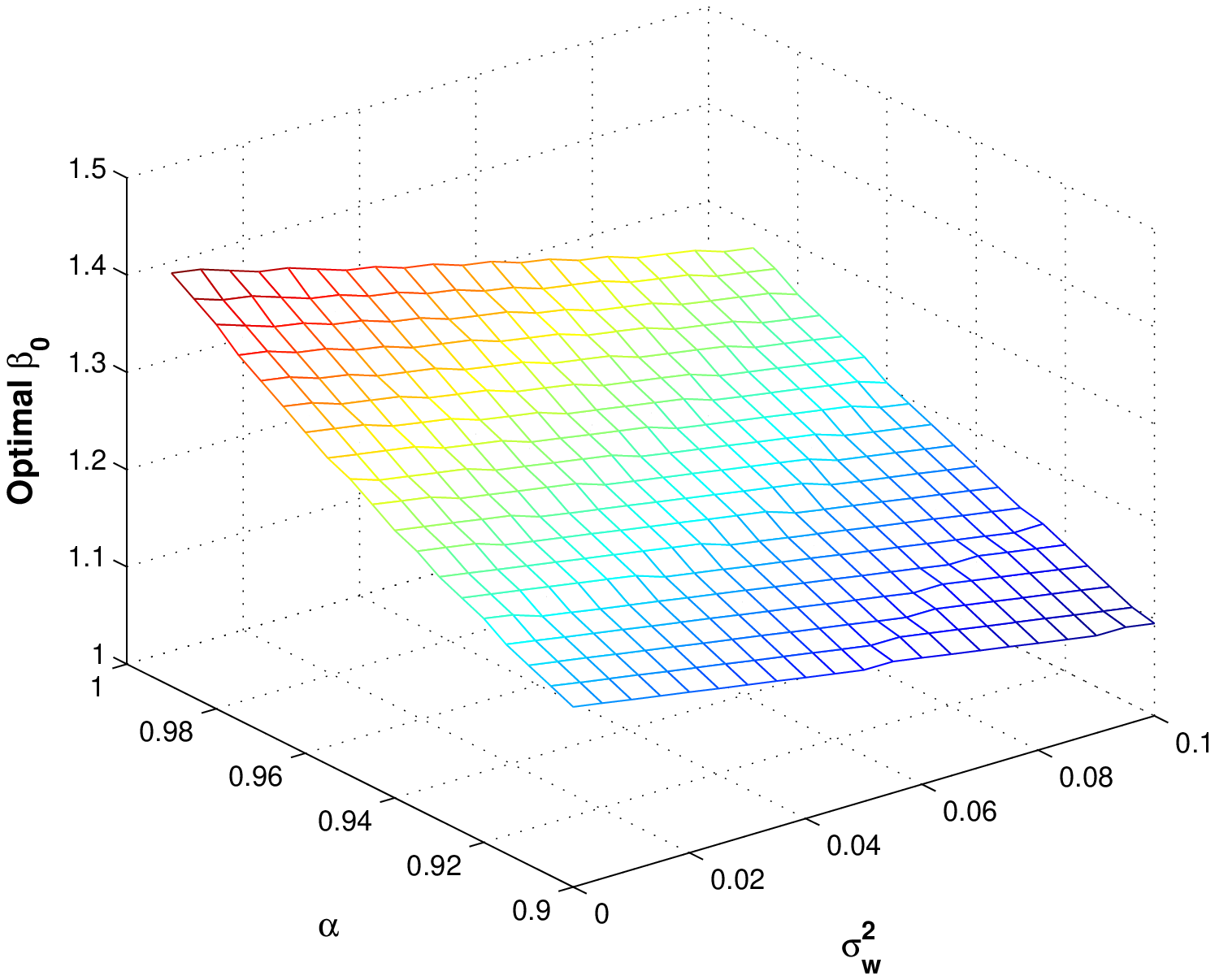}&
    \includegraphics[width=0.5\linewidth]{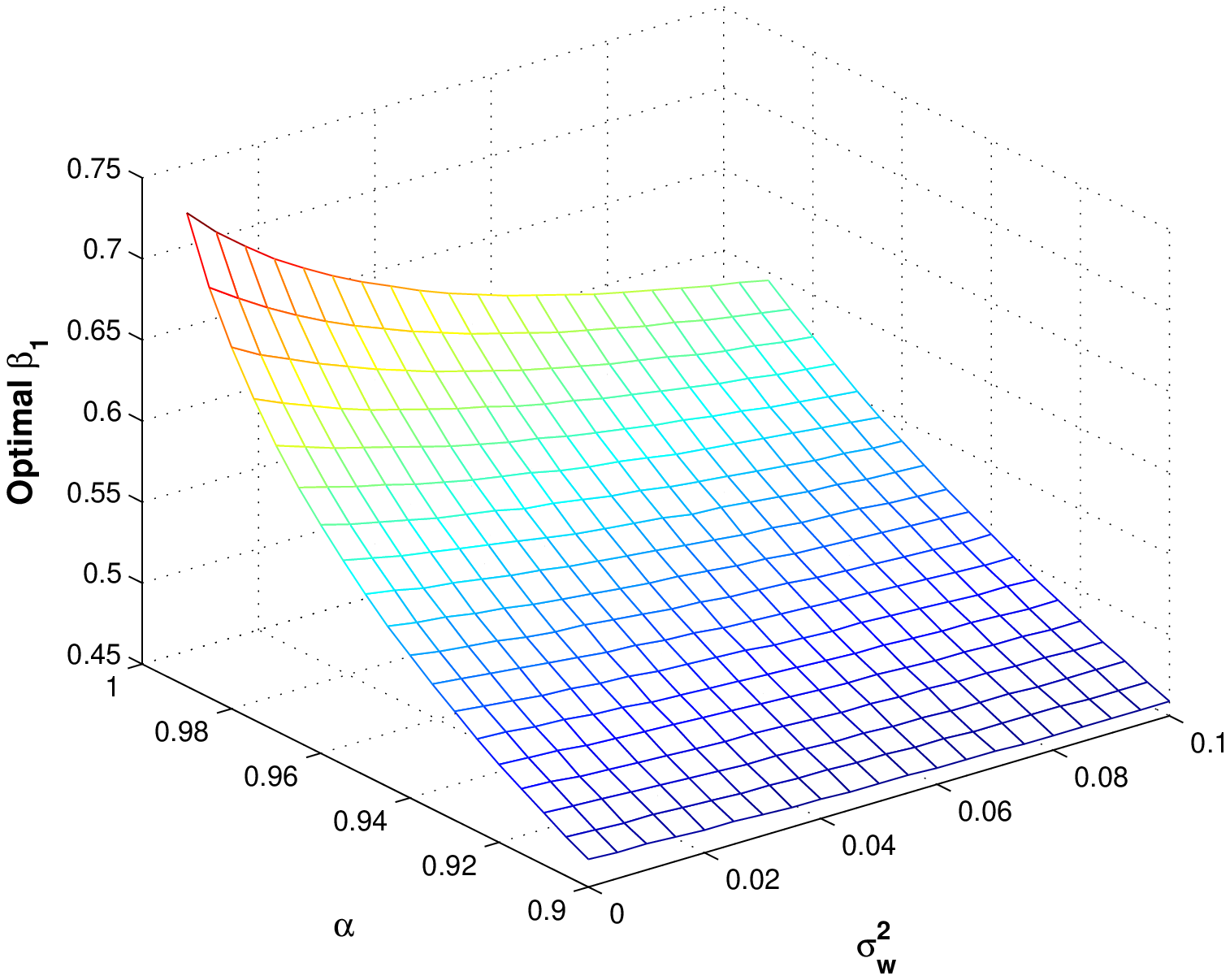}\\
    \scriptsize{(a)}&
    \scriptsize{(b)}\\
\end{tabular}
\caption{The effect of channel estimation error ($\sigma_w^2$) and feedback delay ($\alpha$) on the optimal value of $\beta_0$ and $\beta_1$
($\rho=10$ dB, $K=10$, $\sigma_w^2=0:0.005:0.1$, $\alpha=0.9:0.005:0.99$): (a) The optimal fix rate parameter $\beta_0$ with respect to
$\sigma_w^2$ and $\alpha$; (b) The optimal variable rate parameter $\beta_1$ with respect to $\sigma_w^2$ and $\alpha$.} \label{fig_7}
\end{figure}

\subsection{Imperfect Feedback Scenario}\label{numerical_imperfect}
Fig. \ref{fig_6} exhibits the comparison between the fix rate and variable rate outage scenarios as well as the effect of the number of users on
the optimization of $\beta_0$ and $\beta_1$. In order to show the system performance of the two scenarios in one figure, a normalized parameter
$\beta$ is defined. While examining the variable rate plots $\beta=\beta_1,$ and when considering the fixed rate plots $\beta=\beta_0/10$. The
system parameters are: $\alpha=0.98$, $\sigma_w^2=0.01$, and $\rho=10$ dB. It can be seen that for both scenarios, larger number of users $K$
yields better system performance, i.e., higher average goodput and lower average outage probability. This is a consequence of increased
multiuser diversity gain to combat the imperfections in the feedback system.

\begin{figure}[t]
\centering
    \includegraphics[width=0.6\linewidth]{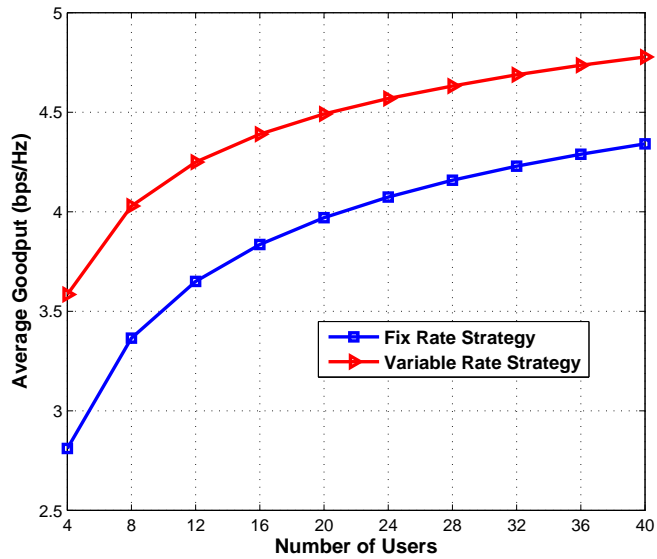}
\caption{Comparison of the average goodput for a $4$-cluster system with fix rate and variable rate strategies using optimized $\beta_0^*$ and
$\beta_1^*$. The average goodput is calculated using the best-M partial feedback scheme when the minimum required $M^*$ is computed after
obtaining $\beta_0^*$ or $\beta_1^*$. ($N=64$, $\boldsymbol\eta=(1,2,4,8)$, $K_1=K_2=K_3=K_4=K/4$, $\alpha=0.98$, $\sigma_w^2=0.01$, $\rho=10$
dB)} \label{fig_8}
\end{figure}

Fig. \ref{fig_7} illustrates the effect of channel estimation error $\sigma_w^2$ and feedback delay $\alpha$ on the optimal value of $\beta_0$
and $\beta_1$. Here $\sigma_w^2$ is varied from $0$ to $0.1$, and $\alpha$ is varied from $0.9$ to $0.99$ in steps of $0.005$. It can be
observed from the changing profiles that both the optimal values of $\beta_0$ and $\beta_1$ get smaller as the imperfections become worse.
Therefore, the system should adjust the system parameters to adapt to the encountered imperfections.

Now we consider the adaptation of system parameters ($\beta_0$ or $\beta_1$) and partial feedback in a $4$-cluster heterogeneous feedback
system. The system parameters are: $\boldsymbol\eta=(1,2,4,8)$, $\alpha=0.98$, $\sigma_w^2=0.01$, and $\rho=10$ dB. For both transmission
strategies and for a given number of users $K$, the optimal value of $\beta_0^*$ or $\beta_1^*$ is first optimized according to the full
feedback case discussed in Section \ref{optimize}. Then, a minimum required $M^*$ is obtained by matching the system performance to that in the
full feedback case. Fig. \ref{fig_8} demonstrates the average goodput for both transmission strategies with $M^*$ and $\beta_0^*$ (or
$\beta_1^*$). We observe that there is almost a constant performance gain for the variable rate strategy compared with the fix rate one. This is
due to the fact that for the variable rate scenario, the system is adapting the transmission parameters conditioned on the past memory even if
it is the outdated one. If the channel estimation error and feedback delay are not severe, the imperfections can be compensated by multiplying
with the backoff factor and relying on the past feedback.

\section{Conclusion}\label{conclusion}
In this paper, we propose and analyze a heterogeneous feedback design adapting the feedback resource according to users' frequency domain
channel statistics. Under the general correlated channel model, we demonstrate the gain by achieving the potential match among coherence
bandwidth, subband size and partial feedback. To facilitate statistical analysis, we employ the subband fading model for the multi-cluster
heterogeneous feedback system. We derive a closed form expression of the average sum rate under perfect partial feedback assumption, and provide
a method to obtain the minimum heterogeneous partial feedback required to obtain performance comparable to a scheme using full feedback. We also
analyze the effect of imperfections on the heterogeneous partial feedback system. We obtain a closed form expression for the average goodput of
the fix rate scenario, and utilize a bounding technique and tight approximation to track the performance of the variable rate scenario. Methods
adapting the system parameters to maximize the average system goodput are proposed. The heterogeneous feedback design is shown to outperform the
homogeneous one with the same feedback resource. With imperfections, the system adjusting the transmission strategy and the amount of partial
feedback is shown to yield better performance. The developed analysis provides a theoretical reference to understand the approximate behavior of
the proposed heterogeneous feedback system and its interplay with practical imperfections. Dealing with the general channel correlation and the
corresponding nonlinear nature of the CQI are interesting directions for the heterogeneous feedback system.

% if have a single appendix:
%\appendix[Proof of the Zonklar Equations]
% or
%\appendix  % for no appendix heading
% do not use \section anymore after \appendix, only \section*
% is possibly needed

% use appendices with more than one appendix
% then use \section to start each appendix
% you must declare a \section before using any
% \subsection or using \label (\appendices by itself
% starts a section numbered zero.)
%

\appendices
\section{}\label{appenA}
\textit{Proof (sketch) of Lemma \ref{lemma_1}:} The methodology is an extension of the work in \cite{hur11} which deals with the homogeneous
feedback case with one cluster of users and one specific subband size.

Let $F_{Y_{k}^{(g)}}$ denote the CDF of $Y_{k}^{(g)}\triangleq Y_{k,r}^{(g)}$. Substituting the subband size $\frac{N}{\eta_g}$ and the number
of reported CQI $\frac{\eta_G}{\eta_g} M$ for user $k$ in cluster $\mathcal{K}_g$ makes $F_{Y_{k}^{(g)}}$ satisfy (\ref{perfect:eq_1}). It can
be shown that $F_{\tilde{Y}_{k}^{(g)}}(x)=\mathbb{P}(\tilde{Y}_{k,n}^{(g)}\leq x)=\mathbb{P}(Y_{k,\lceil\frac{n}{\eta_g}\rceil}^{(g)}\leq
x)=F_{Y_{k}^{(g)}}(x)$, which concludes the proof.

\medskip

\textit{Proof of Theorem \ref{theorem_1}:} Substituting the expressions of $F_{\tilde{Y}_{k}^{(g)}}$ from Lemma \ref{lemma_1} and combining
(\ref{perfect:eq_5}) yield:
\begin{equation} \label{appendix:eq_1}
F_{X_n|\mathcal{U}_n}(x)=(F_Z(x))^{\sum_{g=1}^G\frac{N}{\eta_g}\tau_g}\prod_{g=1}^G\left(\sum_{m=0}^{\frac{\eta_G}{\eta_g}M-1}\frac{\xi_g(N,M,\boldsymbol\eta,0)}{(F_Z(x))^m}\right)^{\tau_g}.
\end{equation}

Applying \cite[0.314]{gradshteyn07} to a finite-order power series in (\ref{appendix:eq_1}),
$\left(\sum_{m=0}^{\frac{\eta_G}{\eta_g}M-1}\frac{\xi_g(N,M,\boldsymbol\eta,0)}{(F_Z(x))^m}\right)^{\tau_g}$ can be expressed as
$\sum_{m=0}^{\tau_g(\frac{\eta_G}{\eta_g}M-1)}\frac{\Lambda_g(N,M,\boldsymbol\eta,\boldsymbol\tau,m)}{(F_Z(x))^m}$, where the expression for
$\Lambda_g(N,M,\boldsymbol\eta,\boldsymbol\tau,m)$ is described in Theorem \ref{theorem_1}. Note that the coefficients of $\frac{1}{(F_Z(x))^m}$
can be computed in a recursive manner.

Then we employ \cite[0.316]{gradshteyn07} for the multiplication of power series. For $g=1$, $\Theta_1(N,M,\boldsymbol\eta,\boldsymbol\tau,m)$
can be calculated from $\Lambda_1(N,M,\boldsymbol\eta,\boldsymbol\tau,m)$ and $\Lambda_2(N,M,\boldsymbol\eta,\boldsymbol\tau,m)$ as:
\begin{equation*}
\Theta_1(N,M,\boldsymbol\eta,\boldsymbol\tau,m)=\mathop{\sum}\limits_{i=0}^m\Lambda_1(N,M,\boldsymbol\eta,\boldsymbol\tau,i)\Lambda_2(N,M,\boldsymbol\eta,\boldsymbol\tau,m-i).
\end{equation*}
For $2\leq g<G$, $\Theta_{g}(N,M,\boldsymbol\eta,\boldsymbol\tau,m)$ can be computed from $\Theta_{g-1}(N,M,\boldsymbol\eta,\boldsymbol\tau,m)$
and $\Lambda_{g+1}(N,M,\boldsymbol\eta,\boldsymbol\tau,m)$ in the following manner:
\begin{equation*}
\Theta_{g}(N,M,\boldsymbol\eta,\boldsymbol\tau,m)=\mathop{\sum}\limits_{i=0}^m\Theta_{g-1}(N,M,\boldsymbol\eta,\boldsymbol\tau,i)\Lambda_{g+1}(N,M,\boldsymbol\eta,\boldsymbol\tau,m-i).
\end{equation*}
This concludes the proof.

\medskip

\textit{Derivation of $\mathcal{I}_1(a,b)$:} From the definition of $Z$, $F_Z(x)=1-\exp(-x)$ and $f_Z(x)=\exp(-x)$. Thus
$d(F_Z(x))^b=b(F_Z(x))^{b-1}f_Z(x)dx=b\sum_{\ell=0}^{b-1}\binom{b-1}{\ell}(-1)^{\ell}\exp(-(\ell+1)x)dx$, where the last equality follows from
the binomial theorem.

Therefore, $\int_0^\infty
\log_2(1+ax)d(F_Z(x))^b=\frac{b}{\ln2}\sum_{\ell=0}^{b-1}\binom{b-1}{\ell}(-1)^{\ell}\int_0^\infty\ln(1+ax)\exp(-(\ell+1)x)dx$. Applying
\cite[4.337.2]{gradshteyn07} yields (\ref{perfect:eq_10}).

% you can choose not to have a title for an appendix
% if you want by leaving the argument blank
\section{}\label{appenB}

\textit{Derivation of $\mathcal{I}_2(a,b)$:} From the definition of $\hat{Z}$, it can be shown that
$F_{\hat{Z}}(x)=1-\exp(-\frac{1}{1-\sigma_w^2}x)$ and $f_{\hat{Z}}(x)=\frac{1}{1-\sigma_w^2}\exp(-\frac{1}{1-\sigma_w^2}x)$. Then
$\mathcal{I}_2(a,b)$ can be calculated as:
\begin{align}
\mathcal{I}_2(a,b)&\mathop{=}\limits^{(a)}\frac{2b}{(1-\sigma_w^2)\ln2}\sum_{\ell=0}^{b-1}\binom{b-1}{\ell}(-1)^{\ell}\int_0^\infty\mathcal{Q}_1(\alpha_w\alpha
x,\alpha_w\sqrt{a})\exp\left(-\frac{(\ell+1)x^2}{1-\sigma_w^2}\right)xdx\notag\\
&\mathop{=}\limits^{(b)}\frac{2b}{(1-\sigma_w^2)\ln2}\sum_{\ell=0}^{b-1}\binom{b-1}{\ell}(-1)^{\ell}\frac{1}{\zeta_{\ell}}\left(\mathcal{Q}_1(0,\vartheta)+\exp(-\frac{\zeta_{\ell}\vartheta^2}{2(\varpi^2+\zeta_{\ell})})(1-\mathcal{Q}_1(0,\frac{\varpi\vartheta}{\sqrt{\varpi^2+\zeta_{\ell}}}))\right)\notag\\
\label{appendix:eq_2}&\mathop{=}\limits^{(c)}\frac{2b}{(1-\sigma_w^2)\ln2}\sum_{\ell=0}^{b-1}\binom{b-1}{\ell}(-1)^{\ell}\frac{1}{\zeta_{\ell}}\left(\exp(-\frac{\vartheta^2}{2})+\exp(-\frac{\zeta_{\ell}\vartheta^2}{2(\varpi^2+\zeta_{\ell})})(1-\exp(-\frac{\varpi^2\vartheta^2}{2(\varpi^2+\zeta_{\ell})}))\right),
\end{align}
where $\varpi=\alpha_w\alpha$, $\vartheta=\alpha_w\sqrt{a}$, $\zeta_{\ell}=\frac{2(\ell+1)}{1-\sigma_w^2}$. (a) is obtained by substituting the
expression of $d(F_{\hat{Z}}(x))^b$ and using change of variables; (b) follows from applying \cite[B.18]{simon02}; (c) follows from using the
fact that $\mathcal{Q}_1(0,\vartheta)=\exp(-\frac{\vartheta^2}{2})$.

\medskip

\textit{Proof of Proposition \ref{proposition_1}:}
\begin{align}
\mathcal{I}_3(a,b)&\mathop{=}\limits^{(a)}\frac{b}{(1-\sigma_w^2)\ln2}\sum_{\ell=0}^{b-1}\binom{b-1}{\ell}(-1)^{\ell}\int_0^\infty\mathcal{Q}_1(\alpha_w\alpha
\sqrt{x},\alpha_w\sqrt{ax})\ln(1+\rho ax)\exp\left(-\frac{(\ell+1)x}{1-\sigma_w^2}\right)dx\notag\\
&\mathop{<}\limits^{(b)}\frac{2\rho
ab}{(1-\sigma_w^2)\ln2}\sum_{\ell=0}^{b-1}\binom{b-1}{\ell}(-1)^{\ell}\int_0^\infty\mathcal{Q}_1(\alpha_w\alpha
x,\alpha_w\sqrt{a}x)\exp\left(-\frac{(\ell+1)x^2}{1-\sigma_w^2}\right)x^3dx\notag\\
&\mathop{=}\limits^{(c)}\frac{4\rho ab}{(1-\sigma_w^2)\ln2}\sum_{\ell=0}^{b-1}(-1)^{\ell}\frac{1}{\zeta_{\ell}^2}\Bigg(1+\frac{\vartheta^2}{\varphi_{\ell}}\Bigg(\frac{\varpi^2}{\varphi_{\ell}}{}_2F_1\left(1,\frac{3}{2};2;\frac{4\varpi^2\vartheta^2}{\varphi_{\ell}^2}\right)-{}_2F_1\left(\frac{1}{2},1;1;\frac{4\varpi^2\vartheta^2}{\varphi_{\ell}^2}\right)\notag\\
\label{appendix:eq_3}&\quad+\frac{2\zeta_{\ell}}{\varphi_{\ell}}\Bigg(\frac{\varpi^2}{\varphi_{\ell}}{}_2F_1\left(\frac{3}{2},2;2;\frac{4\varpi^2\vartheta^2}{\varphi_{\ell}^2}\right)-\frac{1}{2}{}_2F_1\left(1,\frac{3}{2};1;\frac{4\varpi^2\vartheta^2}{\varphi_{\ell}^2}\right)\Bigg)\Bigg)\Bigg),
\end{align}
where $\varpi=\alpha_w\alpha$, $\vartheta=\alpha_w\sqrt{a}$, $\zeta_{\ell}=\frac{2(\ell+1)}{1-\sigma_w^2}$,
$\varphi_{\ell}=\varpi^2+\vartheta^2+\zeta_{\ell}$, and ${}_2F_1(\cdot,\cdot;\cdot;\cdot)$ is the Gaussian hypergeometric function
\cite{abramowitz72}. (a) is obtained by substituting the expression of $d(F_{\hat{Z}}(x))^b$; (b) follows from the fact that when $\rho\ll1$,
$\rho ax$ is a tight upper bound for $\ln(1+\rho ax)$; note that change of variables are used; (c) follows from applying \cite[B.60]{simon02}.

\medskip

\textit{Proof (sketch) of Proposition \ref{proposition_2}:} Define
$s(\check{\chi}_b)\triangleq\mathcal{Q}_1(\alpha_w\alpha\sqrt{\check{\chi}_b},\alpha_w\sqrt{\beta_1\check{\chi}_b})\log_2(1+\rho
\beta_1\check{\chi}_b)]$. Firstly it must be shown that $\frac{s(\check{\chi}_b)}{s(\mathbb{E}[\check{\chi}_b])}$ converges to $1$ in
probability. For $\forall\epsilon>0$, it is now shown that:
\begin{align}
\mathbb{P}\left(\left|\frac{s(\check{\chi}_b)}{s(\mathbb{E}[\check{\chi}_b])}-1\right|\geq\epsilon\right)=&\mathbb{P}\left(\left|\frac{s(\check{\chi}_b)-s(\mathbb{E}[\check{\chi}_b])}{s(\mathbb{E}[\check{\chi}_b])}\right|\geq\epsilon\right)\notag\\
\label{appendix:eq_6}&\mathop{\leq}\limits^{(a)}\mathbb{P}\left(\frac{s(|\check{\chi}_b-\mathbb{E}[\check{\chi}_b]|)}{s(\mathbb{E}[\check{\chi}_b])}\geq\epsilon\right)\mathop{\rightarrow}\limits^{(b)}0,
\end{align}
where (a) follows from the concave and monotonically increasing property of $s(\cdot)$: $|s(x)-s(y)|<s(|x-y|)$; (b) follows from the asymptotic
scaling rate of $\mathbb{E}[\check{\chi}_b]$ and $|\check{\chi}_b-\mathbb{E}[\check{\chi}_b]|$, and the utilization of the Chebyshev's
inequality. From extreme value theory and asymptotic analysis of order statistics \cite{david03, sharif05}, it is known that the tail behavior
of $\check{\chi}_b$ converges to type $3$ Gumbel distribution, which enables $\mathbb{E}[\check{\chi}_b]$ to scale as $\log b$ and
$|\check{\chi}_b-\mathbb{E}[\check{\chi}_b]|$ to scale as $\log\log b$.

Then a method similar to that in \cite{sanayei07} can be employed to prove the uniformly integrable property \cite{billingsley95} of
$\frac{s(\check{\chi}_b)}{s(\mathbb{E}[\check{\chi}_b])}$. By combining the above property along with the convergence in probability leads to
convergence in the mean \cite{billingsley95}, which concludes the proof.

\medskip

\textit{Proof of Proposition \ref{proposition_3}:} It must be shown that
$\mathcal{I}_3^{\mathrm{A}}(\beta_1,K)=\mathcal{Q}_1\left(\alpha_w\alpha\sqrt{\mathbb{E}[\check{\chi}]},\alpha_w\sqrt{\beta_1\mathbb{E}[\check{\chi}]}\right)\log_2(1+\rho
\beta_1\mathbb{E}[\check{\chi}])$ is strictly quasiconcave in $\beta_1$.

This property can be proved by log-concavity \cite{boyd04}. It is shown in \cite{finner97, yu11} that $\mathcal{Q}_1(\sqrt{a},\sqrt{b})$ is
log-concave in $b\in[0,\infty)$ for $a\geq 0$. Also, $\log(1+b)$ is concave thus log-concave in $b\in[0,\infty)$. Since log-concavity is
maintained in multiplication, $\mathcal{Q}_1(\sqrt{a},\sqrt{b})\log(1+b)$ is log-concave in $b\in[0,\infty)$. From the definition of
$\mathcal{I}_3^{\mathrm{A}}(\beta_1,K)$, it is now proved to be log-concave in $\beta_1\in[0,\infty)$ since $\mathbb{E}[\check{\chi}]$ is
irrelevant to $\beta_1$. Therefore, it is quasiconcave in $\beta_1\in[0,\infty)$ because log-concave functions are also quasiconcave.

In addition, it is clear that $\mathop{\lim}\limits_{\beta_1\rightarrow 0}\mathcal{I}_3^{\mathrm{A}}(\beta_1,K)=0$. Also, it is now shown that:
\begin{align}
0&\leq \mathop{\lim}\limits_{\beta_1\rightarrow\infty}\mathcal{I}_3^{\mathrm{A}}(\beta_1,K)\notag\\
&\mathop{\leq}\limits^{(a)}\mathop{\lim}\limits_{\beta_1\rightarrow\infty}\exp\left(-\frac{(\alpha_w\sqrt{\beta_1\mathbb{E}[\check{\chi}]}-\alpha_w\alpha\sqrt{\mathbb{E}[\check{\chi}]})^2}{2}\right)\log_2(1+\rho\beta_1\mathbb{E}[\check{\chi}])\notag\\
\label{appendix:eq_5}&\mathop{=}\limits^{(b)}\mathop{\lim}\limits_{\beta_1\rightarrow\infty}\frac{\rho}{2\alpha_w^2\ln2}\frac{1}{\left(1+\rho\beta_1\mathbb{E}[\check{\chi}]\right)\left(1-\frac{\alpha}{\sqrt{\beta_1}}\right)}\exp\left(-\frac{(\alpha_w\sqrt{\beta_1\mathbb{E}[\check{\chi}]}-\alpha_w\alpha\sqrt{\mathbb{E}[\check{\chi}]})^2}{2}\right)=0,
\end{align}
where (a) follows from the upper bound $\mathcal{Q}_1(a,b)\leq \exp\left(-\frac{(b-a)^2}{2}\right)$ for $b>a\geq0$ \cite{simon02}; (b) follows
from applying L'Hospital's rule. Therefore, there exists a unique global optimal $\beta_1$ which maximizes
$\mathcal{I}_3^{\mathrm{A}}(\beta_1,K)$.

% use section* for acknowledgement
\section*{Acknowledgment}
The authors want to express their deep appreciation to the anonymous reviewers and the Associated Editor for their many valuable comments and
suggestions, which have greatly helped to improve this paper.

% Can use something like this to put references on a page
% by themselves when using endfloat and the captionsoff option.
\ifCLASSOPTIONcaptionsoff
  \newpage
\fi

% trigger a \newpage just before the given reference
% number - used to balance the columns on the last page
% adjust value as needed - may need to be readjusted if
% the document is modified later
%\IEEEtriggeratref{8}
% The "triggered" command can be changed if desired:
%\IEEEtriggercmd{\enlargethispage{-5in}}

% references section

% can use a bibliography generated by BibTeX as a .bbl file
% BibTeX documentation can be easily obtained at:
% http://www.ctan.org/tex-archive/biblio/bibtex/contrib/doc/
% The IEEEtran BibTeX style support page is at:
% http://www.michaelshell.org/tex/ieeetran/bibtex/
%\bibliographystyle{IEEEtran}
% argument is your BibTeX string definitions and bibliography database(s)
%\bibliography{IEEEabrv,../bib/paper}

\begin{thebibliography}{10}

\bibitem{love08}
D.~J. Love, R.~W. Heath, V.~K.~N. Lau, D.~Gesbert, B.~D. Rao, and M.~Andrews,
  ``{An overview of limited feedback in wireless communication systems},''
  \emph{{IEEE} J. Sel. Areas Commun.}, vol.~26, no.~8, pp. 1341--1365, Oct.
  2008.

\bibitem{zhu09}
H.~Zhu and J.~Wang, ``{Chunk-based resource allocation in OFDMA systems-part I:
  chunk allocation},'' \emph{{IEEE} Trans. Commun.}, vol.~57, no.~9, pp.
  2734--2744, Sept. 2009.

\bibitem{asplund06}
H.~Asplund, A.~A. Glazunov, A.~F. Molisch, K.~I. Pedersen, and M.~Steinbauer,
  ``{The COST 259 directional channel model-part II: macrocells},''
  \emph{{IEEE} Trans. Wireless Commun.}, vol.~5, no.~12, pp. 3434--3450, Dec.
  2006.

\bibitem{huang12}
Y.~Huang and B.~D. Rao, ``{Awareness of channel statistics for slow cyclic
  prefix adaptation in an OFDMA system},'' \emph{{IEEE} Wireless Commun.
  Lett.}, vol.~1, no.~4, pp. 332--335, Aug. 2012.

\bibitem{knopp95}
R.~Knopp and P.~A. Humblet, ``Information capacity and power control in
  single-cell multiuser communications,'' in \emph{Proc. IEEE International
  Conference on Communications (ICC)}, Jun. 1995, pp. 331--335.

\bibitem{viswanath02}
P.~Viswanath, D.~N.~C. Tse, and R.~Laroia, ``{Opportunistic beamforming using
  dumb antennas},'' \emph{{IEEE} Trans. Inf. Theory}, vol.~48, no.~6, pp.
  1277--1294, Jun. 2002.

\bibitem{sanayei07}
S.~Sanayei and A.~Nosratinia, ``{Opportunistic downlink transmission with
  limited feedback},'' \emph{{IEEE} Trans. Inf. Theory}, vol.~53, no.~11, pp.
  4363--4372, Nov. 2007.

\bibitem{hassel07}
V.~Hassel, D.~Gesbert, M.~S. Alouini, and G.~E. Oien, ``{A threshold-based
  channel state feedback algorithm for modern cellular systems},'' \emph{{IEEE}
  Trans. Wireless Commun.}, vol.~6, no.~7, pp. 2422--2426, Jul. 2007.

\bibitem{chen08}
J.~Chen, R.~Berry, and M.~Honig, ``{Limited feedback schemes for downlink OFDMA
  based on sub-channel groups},'' \emph{{IEEE} J. Sel. Areas Commun.}, vol.~26,
  no.~8, pp. 1451--1461, Oct. 2008.

\bibitem{pugh10}
M.~Pugh and B.~D. Rao, ``{Reduced feedback schemes using random beamforming in
  MIMO broadcast channels},'' \emph{{IEEE} Trans. Signal Process.}, vol.~58,
  no.~3, pp. 1821--1832, Mar. 2010.

\bibitem{sesia11}
S.~Sesia, I.~Toufik, and M.~Baker, \emph{{LTE--The UMTS Long Term Evolution}},
  2nd~ed.\hskip 1em plus 0.5em minus 0.4em\relax Wiley, 2011.

\bibitem{jung07}
B.~C. Jung, T.~W. Ban, W.~Choi, and D.~K. Sung, ``{Capacity analysis of simple
  and opportunistic feedback schemes in OFDMA systems},'' in \emph{Proc.
  International Symposium on Communications and Information Technologies
  (ISCIT)}, Oct. 2007, pp. 203--208.

\bibitem{ko07}
J.~Y. Ko and Y.~H. Lee, ``{Opportunistic transmission with partial channel
  information in multi-user OFDM wireless systems},'' in \emph{Proc. IEEE
  Wireless Communications and Networking Conference (WCNC)}, Mar. 2007, pp.
  1318--1322.

\bibitem{choi07}
J.~G. Choi and S.~Bahk, ``{Cell-throughput analysis of the proportional fair
  scheduler in the single-cell environment},'' \emph{{IEEE} Trans. Veh.
  Technol.}, vol.~56, no.~2, pp. 766--778, Mar. 2007.

\bibitem{choi08}
Y.~J. Choi and S.~Bahk, ``{Partial channel feedback schemes maximizing overall
  efficiency in wireless networks},'' \emph{{IEEE} Trans. Wireless Commun.},
  vol.~7, no.~4, pp. 1306--1314, Apr. 2008.

\bibitem{pedersen09}
K.~Pedersen, T.~Kolding, I.~Kovacs, G.~Monghal, F.~Frederiksen, and
  P.~Mogensen, ``{Performance analysis of simple channel feedback schemes for a
  practical OFDMA system},'' \emph{{IEEE} Trans. Veh. Technol.}, vol.~58,
  no.~9, pp. 5309--5314, Nov. 2009.

\bibitem{leinonen09}
J.~Leinonen, J.~Hamalainen, and M.~Juntti, ``{Performance analysis of downlink
  OFDMA resource allocation with limited feedback},'' \emph{{IEEE} Trans.
  Wireless Commun.}, vol.~8, no.~6, pp. 2927--2937, Jun. 2009.

\bibitem{donthi11}
S.~Donthi and N.~Mehta, ``{Joint performance analysis of channel quality
  indicator feedback schemes and frequency-domain scheduling for LTE},''
  \emph{{IEEE} Trans. Veh. Technol.}, vol.~60, no.~7, pp. 3096--3109, Sept.
  2011.

\bibitem{hur11}
S.~Hur and B.~D. Rao, ``{Sum rate analysis of a reduced feedback OFDMA downlink
  system employing joint scheduling and diversity},'' \emph{{IEEE} Trans.
  Signal Process.}, vol.~60, no.~2, pp. 862--876, Feb. 2012.

\bibitem{huang11}
Y.~Huang and B.~D. Rao, ``{Environmental-aware heterogeneous partial feedback
  design in a multiuser OFDMA system},'' in \emph{Proc. Asilomar Conference on
  Signals, Systems, and Computers}, Nov. 2011, pp. 970--974.

\bibitem{piantanida09}
P.~Piantanida, G.~Matz, and P.~Duhamel, ``{Outage behavior of discrete
  memoryless channels under channel estimation errors},'' \emph{{IEEE} Trans.
  Inf. Theory}, vol.~55, no.~9, pp. 4221--4239, Sept. 2009.

\bibitem{isukapalli10}
Y.~Isukapalli and B.~D. Rao, ``{Packet error probability of a transmit
  beamforming system with imperfect feedback},'' \emph{{IEEE} Trans. Signal
  Process.}, vol.~58, no.~4, pp. 2298--2314, Apr. 2010.

\bibitem{lau08}
V.~Lau, W.~K. Ng, and D.~S.~W. Hui, ``{Asymptotic tradeoff between cross-layer
  goodput gain and outage diversity in OFDMA systems with slow fading and
  delayed CSIT},'' \emph{{IEEE} Trans. Wireless Commun.}, vol.~7, no.~7, pp.
  2732--2739, Jul. 2008.

\bibitem{wu10}
T.~Wu and V.~Lau, ``{Design and analysis of multi-user SDMA systems with noisy
  limited CSIT feedback},'' \emph{{IEEE} Trans. Wireless Commun.}, vol.~9,
  no.~4, pp. 1446 --1450, Apr. 2010.

\bibitem{akoum10}
S.~Akoum and R.~W. Heath, ``{Limited feedback for temporally correlated MIMO
  channels with other cell interference},'' \emph{{IEEE} Trans. Signal
  Process.}, vol.~58, no.~10, pp. 5219--5232, Oct. 2010.

\bibitem{ma05}
Q.~Ma and C.~Tepedelenlioglu, ``{Practical multiuser diversity with outdated
  channel feedback},'' \emph{{IEEE} Trans. Veh. Technol.}, vol.~54, no.~4, pp.
  1334--1345, Jul. 2005.

\bibitem{kuhne08}
A.~Kuhne and A.~Klein, ``{Throughput analysis of multi-user OFDMA-systems using
  imperfect CQI feedback and diversity techniques},'' \emph{{IEEE} J. Sel.
  Areas Commun.}, vol.~26, no.~8, pp. 1440--1450, Oct. 2008.

\bibitem{dahlman11}
E.~Dahlman, S.~Parkvall, and J.~Skold, \emph{{4G LTE/LTE-Advanced for Mobile
  Broadband}}.\hskip 1em plus 0.5em minus 0.4em\relax Academic Press, 2011.

\bibitem{ericsson03}
Ericsson, ``{System-level evaluation of OFDM - further consideration},'' 3GPP,
  TSG-RAN WG1R1-031303, Tech. Rep., Nov. 2003.

\bibitem{song11}
H.~Song, R.~Kwan, and J.~Zhang, ``{Approximations of EESM effective SNR
  distribution},'' \emph{{IEEE} Trans. Commun.}, vol.~59, no.~2, pp. 603--612,
  Feb. 2011.

\bibitem{donthi11j}
S.~N. Donthi and N.~B. Mehta, ``{An accurate model for EESM and its application
  to analysis of CQI feedback schemes and scheduling in LTE},'' \emph{{IEEE}
  Trans. Wireless Commun.}, vol.~10, no.~10, pp. 3436--3448, Oct. 2011.

\bibitem{wan06}
L.~Wan, S.~Tsai, and M.~Almgren, ``{A fading-insensitive performance metric for
  a unified link quality model},'' in \emph{Proc. IEEE Wireless Communications
  and Networking Conference (WCNC)}, Apr. 2006, pp. 2110--2114.

\bibitem{fan11}
J.~Fan, Q.~Yin, G.~Y. Li, B.~Peng, and X.~Zhu, ``{Adaptive block-level resource
  allocation in OFDMA networks},'' \emph{{IEEE} Trans. Wireless Commun.},
  vol.~10, no.~11, pp. 3966--3972, Nov. 2011.

\bibitem{forney98}
G.~D. Forney~Jr and G.~Ungerboeck, ``{Modulation and coding for linear Gaussian
  channels},'' \emph{{IEEE} Trans. Inf. Theory}, vol.~44, no.~6, pp.
  2384--2415, Oct. 1998.

\bibitem{al96}
N.~Al-Dhahir and J.~M. Cioffi, ``{Optimum finite-length equalization for
  multicarrier transceivers},'' \emph{{IEEE} Trans. Commun.}, vol.~44, no.~1,
  pp. 56--64, Jan. 1996.

\bibitem{weinfurtner02}
S.~H. Muller-Weinfurtner, ``{Coding approaches for multiple antenna
  transmission in fast fading and OFDM},'' \emph{{IEEE} Trans. Signal
  Process.}, vol.~50, no.~10, pp. 2442--2450, Oct. 2002.

\bibitem{mckay08}
M.~R. McKay, P.~J. Smith, H.~A. Suraweera, and I.~B. Collings, ``{On the mutual
  information distribution of OFDM-based spatial multiplexing: exact variance
  and outage approximation},'' \emph{{IEEE} Trans. Inf. Theory}, vol.~54,
  no.~7, pp. 3260--3278, Jul. 2008.

\bibitem{eslami11}
M.~Eslami and W.~A. Krzymien, ``{Net throughput maximization of per-chunk user
  scheduling for MIMO-OFDM downlink},'' \emph{{IEEE} Trans. Veh. Technol.},
  vol.~60, no.~9, pp. 4338--4348, Nov. 2011.

\bibitem{mceliece84}
R.~McEliece and W.~E. Stark, ``{Channels with block interference},''
  \emph{{IEEE} Trans. Inf. Theory}, vol.~30, no.~1, pp. 44--53, Jan. 1984.

\bibitem{medard02}
M.~M{\'e}dard and R.~G. Gallager, ``{Bandwidth scaling for fading multipath
  channels},'' \emph{{IEEE} Trans. Inf. Theory}, vol.~48, no.~4, pp. 840--852,
  Apr. 2002.

\bibitem{david03}
H.~A. David and H.~N. Nagaraja, \emph{{Order Statistics}}, 3rd~ed.\hskip 1em
  plus 0.5em minus 0.4em\relax Wiley-Interscience, 2003.

\bibitem{abramowitz72}
M.~Abramowitz and I.~A. Stegun, \emph{{Handbook of Mathematical Functions with
  Formulas, Graphs, and Mathematical Tables}}.\hskip 1em plus 0.5em minus
  0.4em\relax Dover, 1972.

\bibitem{sharif05}
M.~Sharif and B.~Hassibi, ``{On the capacity of MIMO broadcast channels with
  partial side information},'' \emph{{IEEE} Trans. Inf. Theory}, vol.~51,
  no.~2, pp. 506--522, Feb. 2005.

\bibitem{nuttall72}
A.~H. Nuttall, ``{Some integrals involving the Q-function},'' Naval Underwater
  Systems Center, Tech. Rep., Apr. 1972.

\bibitem{boyd04}
S.~P. Boyd and L.~Vandenberghe, \emph{{Convex Optimization}}.\hskip 1em plus
  0.5em minus 0.4em\relax Cambridge Univ Pr, 2004.

\bibitem{yu11}
Y.~Yu, ``{On log-concavity of the generalized Marcum Q function},'' \emph{Arxiv
  Preprint}, 2011. [Online]. Available: \url{arXiv:1105.5762}

\bibitem{simon02}
M.~K. Simon, \emph{{Probability Distributions Involving Gaussian Random
  Variables: A Handbook For Engineers and Scientists}}.\hskip 1em plus 0.5em
  minus 0.4em\relax Springer Netherlands, 2002.

\bibitem{gradshteyn07}
I.~S. Gradshteyn and I.~M. Ryzhik, \emph{{Tables of Integrals, Series and
  Products}}, 7th~ed., D.~Zwillinger and A.~Jeffrey, Eds.\hskip 1em plus 0.5em
  minus 0.4em\relax Academic Press, 2007.

\bibitem{billingsley95}
P.~Billingsley, \emph{{Probability and Measure}}, 3rd~ed.\hskip 1em plus 0.5em
  minus 0.4em\relax John Wiley \& Sons, 1995.

\bibitem{finner97}
H.~Finner and M.~Roters, ``{Log-concavity and inequalities for Chi-square, F
  and Beta distributions with applications in multiple comparisons},''
  \emph{Statistica Sinica}, vol.~7, pp. 771--788, 1997.

\end{thebibliography}
%
% <OR> manually copy in the resultant .bbl file
% set second argument of \begin to the number of references
% (used to reserve space for the reference number labels box)

%\newpage

% biography section
%
% If you have an EPS/PDF photo (graphicx package needed) extra braces are
% needed around the contents of the optional argument to biography to prevent
% the LaTeX parser from getting confused when it sees the complicated
% \includegraphics command within an optional argument. (You could create
% your own custom macro containing the \includegraphics command to make things
% simpler here.)
%\begin{biography}[{\includegraphics[width=1in,height=1.25in,clip,keepaspectratio]{mshell}}]{Michael Shell}
% or if you just want to reserve a space for a photo:

\begin{IEEEbiography}[{\includegraphics[width=1in,height=1.25in,clip,keepaspectratio]{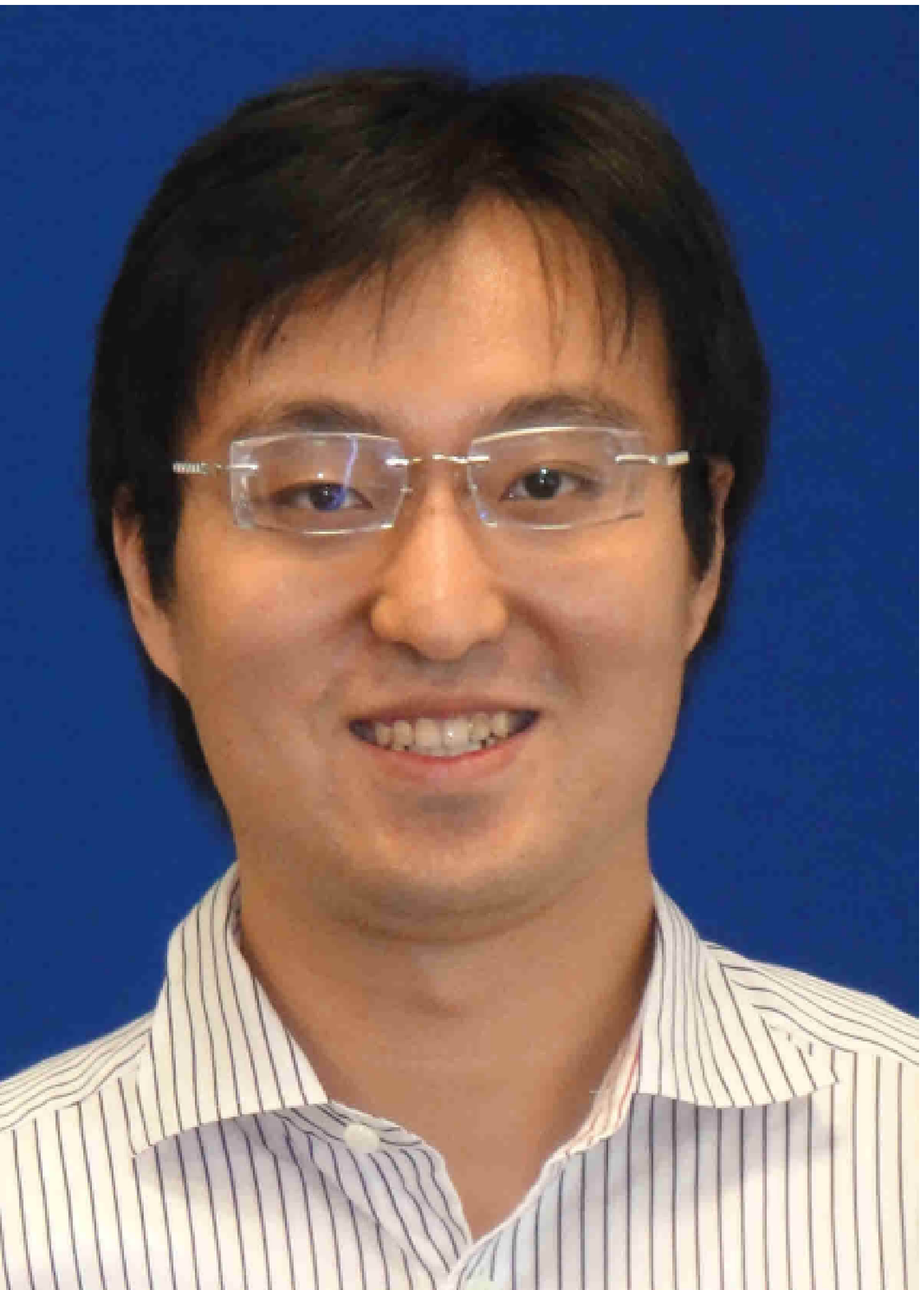}}]{Yichao Huang}
(S'10--M'12) received the B.Eng. degree in information engineering with highest honors from the Southeast University, Nanjing, China, in 2008,
and the M.S. and Ph.D. degrees in electrical engineering from the University of California, San Diego, La Jolla, in 2010 and 2012, respectively.
He then join Qualcomm, Corporate R\&D, San Diego, CA.

He interned with Qualcomm, Corporate R\&D, San Diego, CA, during summer 2011 and summer 2012. He was with California Institute for
Telecommunications and Information Technology (Calit2), San Diego, CA, during summer 2010. He was a visiting student at the Princeton
University, Princeton, NJ, during spring 2012. Mr. Huang received the Microsoft Young Fellow Award in 2007 from Microsoft Research Asia. He
received the ECE Department Fellowship from the University of California, San Diego in 2008, and was a finalist of Qualcomm Innovation
Fellowship in 2010. His research interests include communication theory, optimization theory, wireless networks, and signal processing for
communication systems.
\end{IEEEbiography}
\begin{IEEEbiography}[{\includegraphics[width=1in,height=1.25in,clip,keepaspectratio]{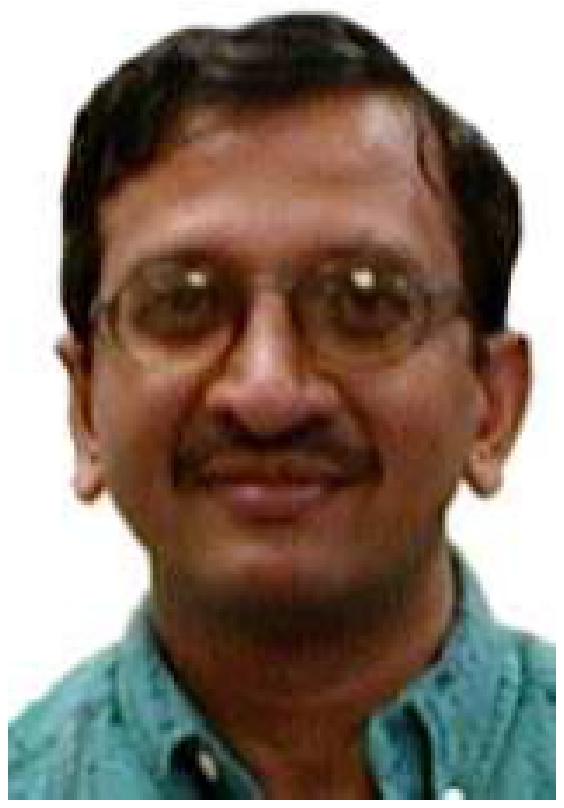}}]{Bhaskar D. Rao}
(S'80--M'83--SM'91--F'00) received the B.Tech. degree in electronics and electrical communication engineering from the Indian Institute of
Technology, Kharagpur, India, in 1979, and the M.Sc. and Ph.D. degrees from the University of Southern California, Los Angeles, in 1981 and
1983, respectively.

Since 1983, he has been with the University of California at San Diego, La Jolla, where he is currently a Professor with the Electrical and
Computer Engineering Department. He is the holder of the Ericsson endowed chair in Wireless Access Networks and was the Director of the Center
for Wireless Communications (2008--2011). His research interests include digital signal processing, estimation theory, and optimization theory,
with applications to digital communications, speech signal processing, and human--computer interactions.

Dr. Rao's research group has received several paper awards. His paper received the Best Paper Award at the 2000 Speech Coding Workshop and his
students have received student paper awards at both the 2005 and 2006 International Conference on Acoustics, Speech, and Signal Processing, as
well as the Best Student Paper Award at NIPS 2006. A paper he coauthored with B. Song and R. Cruz received the 2008 Stephen O. Rice Prize Paper
Award in the Field of Communications Systems. He was elected to the Fellow grade in 2000 for his contributions in high resolution spectral
estimation. He has been a Member of the Statistical Signal and Array Processing technical committee, the Signal Processing Theory and Methods
technical committee, and the Communications technical committee of the IEEE Signal Processing Society. He has also served on the editorial board
of the EURASIP Signal Processing Journal.
\end{IEEEbiography}

% if you will not have a photo at all:
%\begin{IEEEbiographynophoto}{Yichao Huang}
%Biography text here.
%\end{IEEEbiographynophoto}

% insert where needed to balance the two columns on the last page with
% biographies
%\newpage

%\begin{IEEEbiographynophoto}{Bhaskar D. Rao}
%Biography text here.
%\end{IEEEbiographynophoto}

% You can push biographies down or up by placing
% a \vfill before or after them. The appropriate
% use of \vfill depends on what kind of text is
% on the last page and whether or not the columns
% are being equalized.

%\vfill

% Can be used to pull up biographies so that the bottom of the last one
% is flush with the other column.
%\enlargethispage{-5in}

% that's all folks
\end{document}